%% file: Clip3D.tex
\documentclass[10pt,journal]{IEEEtran}
\input{setting-ieee}
\graphicspath{{./figs/}{../}}

\usepackage[round-pad=false, round-mode=places,round-precision=2,group-separator={,},output-decimal-marker={.}]{siunitx}
\usepackage{mathtools}

\begin{document}

\title{
    CLIP-3D: Closed-Loop Evaluation of Performance and Physical Constraints for 3D ICs
}

\author{Shuo~Ren, Libo~Shen, Yaohui~Han, Leilei~Jin, Chenghan~Wang,\\
Zhen~Zhuang, Rongliang~Fu${^*}$,
Bei~Yu,~\IEEEmembership{Senior Member,~IEEE}, Tsung-Yi~Ho,~\IEEEmembership{Fellow,~IEEE}
    \IEEEcompsocitemizethanks{
        \IEEEcompsocthanksitem Shuo~Ren, Libo~Shen, Yaohui~Han, Leilei~Jin, Chenghan~Wang, Zhen~Zhuang, Rongliang~Fu, Bei~Yu, and Tsung-Yi~Ho are with The Chinese University of Hong Kong.
        E-mail: \{sren, lbshen24, yhhan25, lljin, chwang25, zzhuang21, rlfu, ~byu,~tyho\}@cse.cuhk.edu.hk.
        \IEEEcompsocthanksitem  ${^*}$Corresponding author: Rongliang Fu.
    }
}

\maketitle
\pagestyle{plain}

\input{doc/0-abstract}


\input{doc/1-intro}
\input{doc/2-prelim}
\input{doc/3-Methodology}
\input{doc/4-experiment}
\input{doc/5-conclusion}
\input{doc/6-acknowledgments}

\bibliographystyle{IEEEtran}
\bibliography{ref/Top,ref/reference}

\input{doc/bio}

\end{document}

%% file: setting-ieee.tex
\usepackage{blkarray}                                      %
\usepackage{graphicx}                                      %
\usepackage{amsmath}
\usepackage{amssymb}
\usepackage{amsfonts}
\usepackage{amsthm}
\usepackage[cmintegrals]{newtxmath}
\usepackage[mathcal]{eucal}
\usepackage{mathrsfs}
\usepackage{booktabs}
\usepackage{enumerate}
\usepackage{multirow}
\usepackage[subrefformat=parens,farskip=0pt,justification=centering]{subfig}
\usepackage{color}
\usepackage{cite}                                          %
\usepackage{comment}                                       %
\usepackage{soul}                                          %
\soulregister\cite7
\soulregister\ref7
\soulregister\pageref7
\usepackage{etoolbox}                                      %
\usepackage{url}
\usepackage{nth}                                           %
\usepackage{bm}                                            %
\usepackage{courier}
\usepackage{balance}
\usepackage{threeparttable}
\usepackage{xcolor,colortbl}
\usepackage{footnote}
\usepackage{listings}
\usepackage{setspace}                                      %
\usepackage[inline]{enumitem}
\usepackage{verbatim}
\usepackage[bookmarks=false]{hyperref}
\hypersetup{
    colorlinks = true,
    citecolor  = blue,
    linkcolor  = blue,
    urlcolor   = blue,
}
\usepackage{tikz}
\usetikzlibrary{patterns,snakes}
\usetikzlibrary{positioning,calc,fit,decorations.pathmorphing,shapes.geometric, shapes.gates.logic.US, calc}
\usetikzlibrary{arrows,arrows.meta,decorations.markings,shapes,shapes.arrows}
\usetikzlibrary{decorations,decorations.pathreplacing}
\usetikzlibrary{backgrounds}
\usepackage{filecontents}                                  %
\usepackage{pgfplots}
\usepackage{pgfplotstable}
\usepackage{scalefnt}
\pgfplotsset{compat=newest}
\usepackage{caption}
\usepackage{pifont}                                        %
\usepackage{cleveref}
\Crefformat{figure}{Fig.~#2#1#3}                           %
\Crefname{subfigure}{Fig.}{Figs.}
\Crefname{figure}{Fig.}{Figs.}
\Crefformat{table}{TABLE~#2#1#3}                           %
\usepackage[figuresright]{rotating}

\usepackage[linesnumbered,ruled,vlined]{algorithm2e}
\usepackage{algpseudocode}

\definecolor{CUHKorange}{RGB}{244,106,18} %
\definecolor{CUHKblue}{RGB}{0,111,190}    %
\definecolor{CUHKgreen}{RGB}{0,127,128}   %
\definecolor{CUHKred}{RGB}{228,46,36}     %
\definecolor{CUHKyellow}{RGB}{198,148,34} %
\definecolor{CUHKdark}{RGB}{114,44,114}   %
\definecolor{CUHKmiddle}{RGB}{144,44,144} %
\definecolor{CUHKlight}{RGB}{167,44,167} 
\definecolor{CUHKpurple}{RGB}{117,15,109}
\definecolor{CUHKgold}{RGB}{221,163,0}
\definecolor{CUHKribbon}{RGB}{244,223,176}
\definecolor{CUHKblack}{RGB}{34,24,21}

\renewcommand{\bf}[1]{\textbf{#1}}
\renewcommand{\tt}[1]{\texttt{#1}}

\usepackage{tcolorbox}
\tcbuselibrary{skins,breakable}
    {\endtcolorbox}
    {\endtcolorbox}

\crefname{mytheorem}{Theorem}{Theorems}
\crefname{mylemma}{Lemma}{Lemmas}
\crefname{myclaim}{Claim}{Claims}
\crefname{myproperty}{Property}{Properties}
\crefname{mycorollary}{Corollary}{Corollaries}
\crefname{algocf}{algorithm}{algorithms}
\Crefname{algocf}{Algorithm}{Algorithms}

\RequirePackage[normalem]{ulem} %
\RequirePackage{color}\definecolor{RED}{rgb}{1,0,0}\definecolor{BLUE}{rgb}{0,0,1} %

%% file: doc/0-abstract.tex
\begin{abstract}
3D integration packs more power into a smaller footprint, so a candidate design's actual throughput depends on its layout: which macro sits on which tier, where the hot spot lands, and how cache geometry maps to access cycles.
Architectural simulators like gem5 report IPC under idealized timing.
They do not produce the per-block power map, the cache cycle counts, or the 3D layout that decide the realized billion-instructions-per-second (BIPS), so early-stage 3D-IC exploration selects designs without accounting for the effects that decide whether they throttle on silicon.
We present CLIP-3D, a shift-left flow that exposes 3D layout-driven thermal, wire, and cache effects to early-stage architectural exploration before any sign-off tool is invoked.
The first stage lifts an architectural configuration into a physical block representation: McPAT for per-block dynamic and leakage power, CACTI for cache geometry and access cycles, and a HotSpot-compatible 3D stack discretization.
The second stage runs an analytical 3D thermal-aware floorplanner over that representation.
The floorplanner objective embeds a closed-form sustained-frequency expression derived from the linearity of HotSpot's steady-state operator and the standard CMOS power-frequency decomposition. Cross-tier macro assignment and in-plane placement are co-optimized for the realized BIPS rather than for a half-perimeter wirelength (HPWL)-plus-temperature surrogate with hand-tuned weights.
On an 80-configuration architecture sweep under stressed cooling, the floorplanner delivers a realized BIPS gain of $13.75\%$ on average (max $24.77\%$) over the shelf-pack baseline on every one of the 48 thermally-binding configurations and a wire-driven $5.82\%$ on average on the 32 headroom configurations, when each method's optimized layout is back-annotated into gem5 to capture layout-conditional cache and wire latency.
The same realized BIPS plateau is reached by a cool3d-canonical baseline and an SA-plus-$\lambda$-mix grid-search baseline at $2.0\times$ and $2.9\times$ more wall time respectively, so in these experiments CLIP-3D lies on the Pareto frontier of wall time versus realized BIPS.
\end{abstract}

%% file: doc/1-intro.tex
\section{Introduction}
\label{sec:intro}

Modern workloads keep demanding more compute than a single die can provide, and 3D integration is one architectural response: multiple active tiers stack vertically so on-chip compute and cache density grow without enlarging the footprint~\cite{xing2024codesign3DIC}.

Stacking has a cost.
Power dissipates into a smaller volume, so a 3D stack heats up faster than a planar die.
Once the peak die temperature $T_{\max}$ crosses a safe threshold, the chip throttles its clock to protect itself.
The useful throughput is no longer the nominal $\text{IPC}\cdot f_0$ that a simulator reports, but the thermally-derated $\text{IPC}\cdot f(T_{\max})$ delivered on silicon, which is the BIPS metric~\cite{huang2024evaluation}.
What $f(T_{\max})$ ends up being on a given configuration depends on the joint power map across tiers, on which macro sits above which, and on where the hottest concentration of switching activity lands in plane.
The 3D layout decides all of this, and an architectural simulator sees none of it before architecture commitment.

\begin{figure}[t!]
    \centering
    \includegraphics[width=\linewidth]{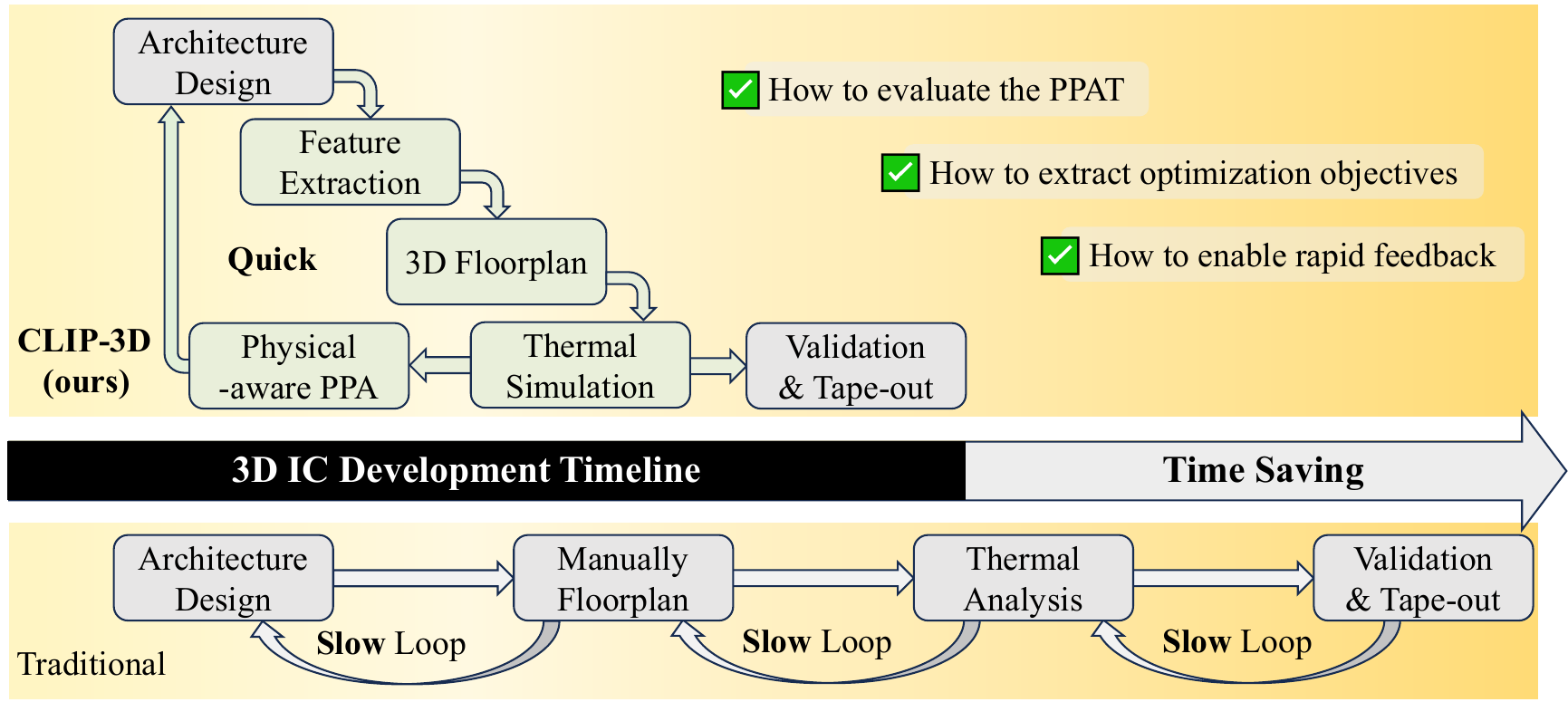}
    \caption{Two existing 3D-IC exploration regimes and where CLIP-3D fits. The Quick path runs an architectural simulator only and reports power, performance, and area (PPA) without thermal. The Traditional path adds McPAT, CACTI, and HotSpot to produce PPA plus thermal (PPAT), but at much higher wall time per candidate.}
    \label{fig:intro_fig}
\end{figure}
Yet early-stage 3D-IC exploration still picks candidates from the simulator's IPC alone.
Simulators such as gem5~\cite{gem5_simulator, gem5_simulator_v20} and SimpleScalar~\cite{SimpleScalar_simulator} assume idealized interconnect latencies and a fixed clock.
The architect sees an IPC per candidate but no per-block power map, no cache cycle counts derived from real array geometry, and no layout under which $T_{\max}$ could be defined.
This choice can fail in a 3D stack: a design that maximizes IPC under ideal timing can also be the design that concentrates the most power in the hottest tier, throttles the most, and delivers among the lowest realized BIPS.

This is a shift-left problem in 3D-IC CAD: the layout-driven physical effects that determine realized BIPS must be exposed to early-stage architectural exploration, not deferred until post-floorplan sign-off.
What is missing is a path that turns the simulator's logic-level output into a physical block representation, plus an optimizer that uses that representation to choose a layout instead of inheriting whatever default the thermal sign-off tool reads from disk.

Two prior threads each address half of this gap.
3D thermal-aware floorplanners~\cite{2004ICCAD_JasonCong_t3dflp,2006_JasonCong_theraml_PD,2006ASPDAC_JasonCong_meva3d,2007ICCD_JasonCong_3d_exploration,2021TVLSI-thermal-aware-3dfloorplan-tsv,sa3dfp_guan_2023} optimize a linear cost $w_{\text{HPWL}}\!\cdot\!\text{HPWL}+w_{\text{thermal}}\!\cdot\!T_{\max}$ (half-perimeter wirelength, HPWL) or a close variant.
The weights are not derived from the architectural quantity the system delivers, so the chosen layout is optimal for a surrogate, not for realized BIPS, and the resulting frequency constraints are never fed back to the architectural simulator.
3D thermal frameworks such as Cool-3D~\cite{wang2025cool3d} and HotGauge~\cite{hankin2021hotgauge}, and runtime schedulers such as HotLEGO~\cite{HotLEGO}, focus on thermal management with the floorplan held fixed: they manage temperature through dynamic voltage and frequency scaling (DVFS) or task migration on a layout the flow takes as input.
Designers therefore still lack a way to answer the concrete question of what BIPS a given microarchitecture and stack deliver under their BIPS-maximizing layout, which depends on physical placement that the simulator never exposes.

To solve these problems,
we present CLIP-3D, summarized in~\Cref{fig:intro_fig}. The figure frames the gap as two existing regimes. The Quick regime evaluates a candidate using fast architectural simulation alone (gem5~\cite{gem5_simulator,gem5_simulator_v20}, SimpleScalar~\cite{SimpleScalar_simulator}) and reports PPA without thermal. The Traditional regime invokes the full McPAT~\cite{li2009mcpat}, CACTI~\cite{muralimanohar2007cacti}, and HotSpot~\cite{hotspot_stan2003, hotspot6, hotspot7} stack to produce PPAT, but is orders of magnitude slower per candidate, so it cannot drive a design-space exploration. CLIP-3D answers the three open questions raised by the figure. It lifts each architectural candidate into the PPAT inputs a 3D layout optimizer needs (per-block power map, cache access cycles, and 3D thermal-stack conductance) with one gem5 pass and one HotSpot solve per candidate, evaluates the layout objective in closed form rather than by repeated thermal simulation, and feeds the resulting realized BIPS back to architectural exploration before commitment.
Our main contributions are summarized as follows:
\begin{itemize}
    \item We bridge architectural simulation and 3D layout with a lifting pipeline. It converts a simulator's IPC trace into the per-block power, cache access cycles, and 3D thermal-stack conductance a layout optimizer needs, with no changes to the upstream simulator (\Cref{sec:method}).
    \item We derive a closed-form expression for the thermally sustainable clock. It is exact under HotSpot's steady-state model and replaces the temperature-frequency fixed-point iteration that thermal-aware evaluators normally run (\Cref{subsec:closedform_freq}).
    \item We build an analytical 3D thermal-aware floorplanner that optimizes realized BIPS directly. The objective needs no HPWL-plus-temperature surrogate weights and no in-loop HotSpot calls, both of which gate prior frameworks (\Cref{sec:analytical-placer}).
    \item We evaluate CLIP-3D on five workloads under two cooling envelopes. It sits on the wall-time vs.\ realized-BIPS Pareto frontier against three baselines, and a ranking diagnostic shows that IPC-only selection becomes roughly uncorrelated with realized BIPS once cooling binds (\Cref{sec:exp}).
\end{itemize}

%% file: doc/2-prelim.tex
\section{Preliminaries}
\label{sec:prelim}

This section walks through the lifting pipeline CLIP-3D builds on, the 3D IC structure that makes layout a first-order architectural decision, and the design-selection problem we solve via the two-stage formulation of~\Cref{subsec:formulation}.

\subsection{Architecture-to-Physical Lifting Pipeline}
\label{subsec:flow}

CLIP-3D drives four open-source EDA tools: gem5~\cite{gem5_simulator, gem5_simulator_v20}, McPAT~\cite{li2009mcpat}, HotSpot~\cite{hotspot_stan2003}, and CACTI~\cite{muralimanohar2007cacti}.
The pipeline turns a candidate architecture into the block-level physical inputs a 3D layout optimizer needs.

A candidate design is $\mathbf{d}=(\mathbf{a},\mathbf{p})$.
$\mathbf{a}$ is the architectural vector (core count, L1D/L2 sizes, associativity, pipeline parameters).
$\mathbf{p}$ is the physical vector (die area, per-block tier assignment, in-plane positions).
For workload $\mathbf{w}$, the lifting starts with a gem5 run that uses an out-of-order core under idealized cache and interconnect latencies.
This first run reports a nominal IPC, denoted $\text{IPC}_1$, and per-block activity statistics (cache accesses, misses, branch events, write-backs) that drive every subsequent stage.

McPAT consumes those statistics and returns per-block dynamic power $P_{\text{dyn},b}$, leakage power $P_{\text{leak},b}$, and an architecture-derived block area $A_b$.
In parallel, CACTI characterizes the cache hierarchy and converts L1D/L2 geometry into access cycles.
Combined with a discrete through-silicon-via (TSV) hop model for inter-tier wiring (\Cref{sec:method}), the CACTI cycles form a latency vector $\mathbf{L}(\mathbf{d})$ that captures both on-die cache timing and cross-tier communication cost.
Back-annotating $\mathbf{L}(\mathbf{d})$ into the cache hierarchy and rerunning gem5 returns a calibrated IPC, denoted $\text{IPC}_2$, that reflects the actual memory-system delay.

The per-block power map is rasterized onto the 3D thermal stack defined by $\mathbf{p}$ and handed to HotSpot, which solves the steady-state heat equation and returns the peak die temperature $T_{\max}(\mathbf{d};\mathbf{w})$.
Once $T_{\max}$ exceeds a safe operating threshold $T_{\text{safe}}$, the hardware throttles its clock to avoid thermal damage.
We model this throttle with the standard piecewise-linear form
\begin{equation}
    f(T)=\max\!\Bigl(f_{\min},\; f_0\bigl[1-\alpha\,\max\!\bigl(0,\;T-T_{\text{safe}}\bigr)\bigr]\Bigr),
    \label{eq:throttle}
\end{equation}
with nominal clock $f_0$, throttling slope $\alpha$, and floor frequency $f_{\min}$; specific parameter values are stated in~\Cref{subsec:setup}.
\Cref{subsec:closedform_freq} shows that the linearity of HotSpot's steady-state operator and the CMOS power-frequency decomposition collapse $f(T_{\max})$ into a closed-form expression in $\mathbf{a}$ and $\mathbf{p}$, which lets the floorplanner of~\Cref{sec:analytical-placer} optimize layout for realized BIPS without invoking HotSpot inside its inner loop.

The realized throughput of a design is the thermally-derated BIPS metric~\cite{huang2024evaluation},
\begin{equation}
    \text{BIPS}_2(\mathbf{d};\mathbf{w}) \;=\; \text{IPC}_2(\mathbf{d};\mathbf{w}) \cdot f\!\bigl(T_{\max}(\mathbf{d};\mathbf{w})\bigr),
    \label{eq:bips}
\end{equation}
which both stages of CLIP-3D compute and optimize.

\subsection{3D IC Structure}
\label{subsec:3dic_struct}

CLIP-3D targets monolithic or TSV-based 3D stacks with $N_t$ active device tiers stacked vertically between a bottom substrate and a top heat sink.
Three physical features distinguish this setting from a planar die.

First, vertical interconnects are discrete.
Cross-tier signals go through TSVs whose delay, area, and parasitics differ materially from intra-tier wires~\cite{ahmed2016tsv}, and the number of TSV hops between two blocks depends on their $(\text{tier},x,y)$ assignment.
A latency model that ignores hop discreteness and uses a continuous linear-distance proxy systematically misestimates cross-tier access time.

Second, vertical thermal coupling is asymmetric.
Heat dissipates primarily through the top heat sink, so lower tiers accumulate heat from all tiers above them.
The peak temperature on the hottest tier therefore depends on the joint power map across tiers, not on any single tier in isolation~\cite{2022ISLPED_lim_3dic_ppa_thermal}, and is decided as much by which macros sit on which tier as by how much power each macro draws.

Third, tier-aware floorplanning changes the design variable.
The physical vector $\mathbf{p}$ assigns each microarchitectural block $b\!\in\!B$ to a $(\text{tier},x,y)$ position under non-overlapping, die-area, and tier-consistency constraints.
This is a richer design space than a 2D floorplan, and a different choice of $\mathbf{p}$ shifts $T_{\max}$ and the realized BIPS even at fixed $\mathbf{a}$.

These three features together mean the map $\mathbf{p}\!\to\!(\mathbf{L},T_{\max})$ is neither separable across tiers nor well-approximated by linear-distance proxies.
A meaningful evaluation of architectural choice $\mathbf{a}$ has to know the layout $\mathbf{p}$ that choice will be deployed under, which motivates the two-stage formulation that the next subsection makes precise.

\subsection{Problem Formulation}
\label{subsec:formulation}

\Cref{tab:notation} collects the symbols introduced in \Cref{subsec:flow,subsec:3dic_struct} for reference throughout the paper.

\input{tables/notation}

For workload $\mathbf{w}$ and an admissible architectural set $\mathcal{A}$, CLIP-3D selects the architecture whose realized BIPS, evaluated under the layout that maximizes that BIPS, is highest:
\begin{equation}
    \mathbf{a}^{\star}(\mathbf{w}) \;=\; \arg\max_{\mathbf{a}\in\mathcal{A}}\;
    \underbrace{\max_{\mathbf{p}\in\mathcal{P}(\mathbf{a})}\; \text{IPC}_2(\mathbf{a},\mathbf{p};\mathbf{w})\cdot f\!\bigl(T_{\max}(\mathbf{a},\mathbf{p};\mathbf{w})\bigr)}_{\mathbf{p}^{\star}(\mathbf{a};\mathbf{w}),\ \text{the inner layout problem solved by~\Cref{sec:analytical-placer}}},
    \label{eq:objective}
\end{equation}
subject to a hard thermal bound $T_{\max}(\mathbf{a},\mathbf{p};\mathbf{w})\!\le\!T_{\text{hard}}$ at the junction reliability limit and physical feasibility of the 3D floorplan (non-overlapping blocks, die-area bound, tier-assignment consistency captured by $\mathcal{P}(\mathbf{a})$).
The inner $\max_\mathbf{p}$ defines the layout problem CLIP-3D's analytical floorplanner solves for each candidate architecture.
The outer $\arg\max_\mathbf{a}$ is the architectural-DSE loop that consumes the floorplanner's output.
A baseline shelf-pack layout, used in our comparisons, is recovered as the special case where $\mathbf{p}$ is fixed by deterministic rule rather than optimized against the realized BIPS.

To quantify how much layout optimization changes the architectural ranking, we additionally report a three-level diagnostic over the same design set:
\begin{itemize}
    \item Level-0 (IPC-only): rank designs by $\text{IPC}_1$ alone. Lifting is ignored entirely.
    \item Level-1 (thermal-only): rank designs by $\text{IPC}_1 \cdot f(T_{\max})$. Only the thermal output of the lifting is used.
    \item Level-2 (lifted): rank designs by $\text{IPC}_2 \cdot f(T_{\max})$. The full lifting output is used.
\end{itemize}
The Kendall-$\tau$~\cite{Kendall_s_tau} between any pair of these rankings, $\tau=(C-D)/\binom{n}{2}$ with $C$ and $D$ counting concordant and discordant design pairs, quantifies how much ranking signal is forfeited when a stage of the pipeline is dropped.
We use this diagnostic in~\Cref{sec:exp} to motivate why both lifting and the layout optimization on top of it are needed.

%% file: tables/notation.tex
\begin{table}[t]
\centering
\caption{Notation used in the methodology.}
\label{tab:notation}
\footnotesize
\setlength{\tabcolsep}{3.8pt}
\resizebox{\linewidth}{!}{%
\begin{tabular}{l l}
\toprule
\textbf{Symbol} & \textbf{Meaning} \\
\midrule
$\mathbf{d}=(\mathbf{a},\mathbf{p})$ & Candidate design: architecture $\mathbf{a}$ and physical $\mathbf{p}$ \\
$\mathbf{a}$ & $(N_{\text{cores}},s_{L1D},s_{L2},\text{assoc},\ldots)$ \\
$\mathbf{p}$ & Die size, tier assignment $\{z_i\}$, floorplan $\{(x_i,y_i,w_i,h_i)\}$ \\
$\mathbf{w}\in\mathcal{W}$ & Workload drawn from the five-kernel suite \\
$B$ & Set of McPAT microarchitectural blocks (cores, caches, NoC, MCs) \\
$P_i,\,A_i$ & Per-block steady-state power / area for block $i\!\in\!B$ \\
$\text{IPC}_1$ & First-pass IPC using nominal cache/hop latencies \\
$\text{IPC}_2$ & Calibrated IPC using $\mathbf{L}(\mathbf{d})$ from~\Cref{eq:latency_model} \\
$T_{\max}$ & Steady-state peak temperature from HotSpot on the grid of~\Cref{eq:gridmap} \\
$f_0,\,f_{\min},\,\alpha,\,T_{\text{safe}}$ & Throttling parameters of~\Cref{eq:throttle} \\
$\ell_{\text{CACTI}}$ & CACTI-derived cache access time in cycles (\Cref{tab:cacti}) \\
$h_{\text{TSV}}$ & Integer TSV hop count on critical L1$\rightarrow$L2 path \\
$\mathbf{L}(\mathbf{d})$ & Discrete latency vector back-annotated into Stage-3 gem5 \\
$\pi_0,\pi_1,\pi_2$ & Level-0/1/2 rankings (\Cref{subsec:formulation}) \\
$\tau(\pi_a,\pi_b)$ & Kendall rank correlation $\tau{=}(C{-}D)/\binom{n}{2}$ \\
$G$ & HotSpot grid resolution per tier (default $32$) \\
\bottomrule
\end{tabular}
}
\vspace{-0.5em}
\end{table}

%% file: doc/3-Methodology.tex
\section{Methodology}
\label{sec:method}
\begin{figure*}[t!]
    \centering
    \includegraphics[width=.99\linewidth]{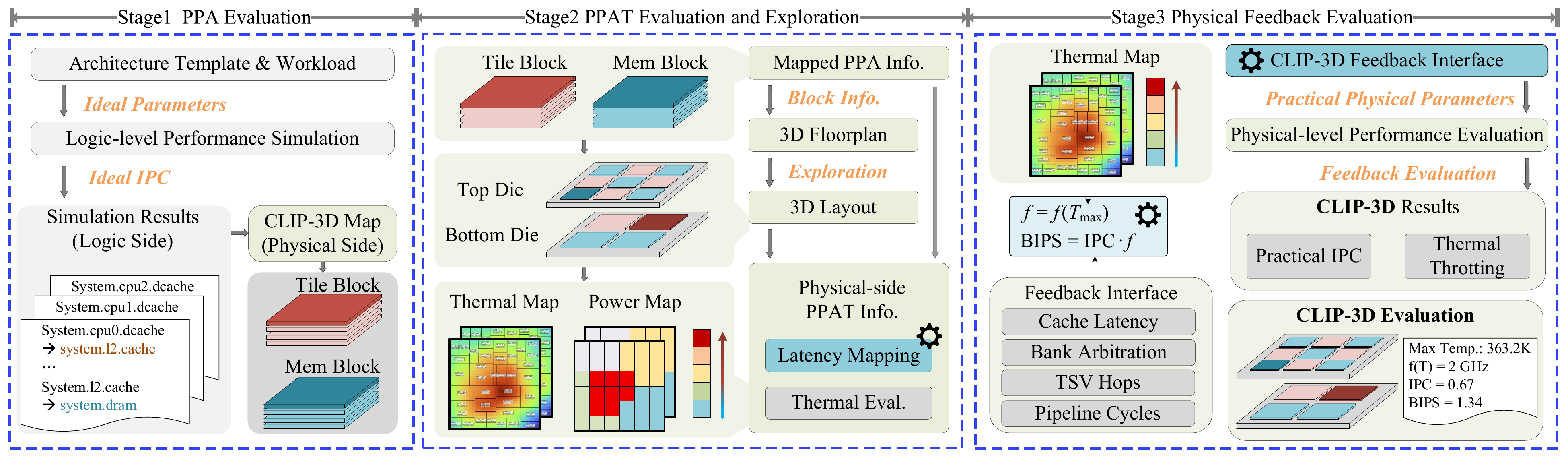}
    \caption{CLIP-3D lifting pipeline: block-level input extraction (Stage~1, left), analytical floorplanning with closed-form $f(T_{\max})$ in the objective (Stage~2, middle), and gem5 back-annotation for the physically feasible $\text{IPC}_2$ (Stage~3, right).}
    \label{fig:overview}
\end{figure*}

CLIP-3D solves the two-stage problem of~\Cref{eq:objective} in three stages (\Cref{fig:overview}).
First, an architecture-to-physical lifting pipeline turns each candidate $\mathbf{a}$ into the block-level inputs (per-block dynamic and leakage power, cache access cycles, 3D thermal-stack conductance) that a 3D layout optimizer needs (\Cref{subsec:overview,subsec:thermal_loop,subsec:delay_loop}).
Second, the linearity of HotSpot's steady-state operator together with the standard CMOS power-frequency decomposition collapse the throttled clock $f(T_{\max})$ into a closed-form expression in those block-level inputs (\Cref{subsec:closedform_freq}), which can be evaluated without re-invoking HotSpot.
Third, an analytical 3D thermal-aware floorplanner consumes the lifted block representation and embeds the closed form directly inside its objective, so the inner $\mathbf{p}^{\star}(\mathbf{a})$ of~\Cref{eq:objective} is solved for the same realized BIPS the rest of the pipeline scores designs on (\Cref{sec:analytical-placer}); symbols follow~\Cref{tab:notation} throughout.

\subsection{Overview and Per-Configuration Lifting}
\label{subsec:overview}
The lifting stage decomposes a candidate $\mathbf{a}$ into the two block-level inputs that decide its realized BIPS: a per-block power map plus 3D-stack thermal model that determines $T_{\max}$ and therefore $f(T_{\max})$, and a CACTI-calibrated cache-and-topology latency vector that is back-annotated into a second gem5 pass for a physically realistic $\text{IPC}_2$.
\begin{itemize}
    \item The thermal lifting (\Cref{subsec:thermal_loop}) maps per-block power and the candidate's stack geometry onto a 3D-stack power grid that HotSpot~\cite{hotspot_stan2003} can solve for the peak die temperature $T_{\max}(\mathbf{d})$, which drives $f(T_{\max})$ via the closed-form model of~\Cref{subsec:closedform_freq}.
    \item The latency lifting (\Cref{subsec:delay_loop}) maps the architectural configuration into a discrete latency vector $\mathbf{L}(\mathbf{d})$: CACTI-calibrated cache access time plus a shared-cache arbitration term, an integer TSV-hop term, and a pipeline constant. $\mathbf{L}(\mathbf{d})$ is back-annotated into a second architectural simulation for a calibrated $\text{IPC}_2$.
\end{itemize}
The two lifted outputs are methodologically disjoint: the floorplan modulates temperature only, and the latency vector depends on architectural and topological quantities only.
This separation is what lets CLIP-3D replace the linear-distance wire-delay proxies of prior flows with a parameter-free discrete model, and it lets the floorplanner of~\Cref{sec:analytical-placer} treat $\mathbf{L}$ as a per-architecture constant while it optimizes layout for $T_{\max}$.

\paragraph{Shared front-end: architecture $\to$ power and $\text{IPC}_1$}
Both lifted outputs start from a single upstream pass.
We run gem5~\cite{gem5_simulator} in syscall-emulation mode with an out-of-order X86 core ($4$-wide dispatch, $192$-entry ROB, split L1 I/D) clocked at $f_0{=}\SI{2.0}{\giga\hertz}$, with the cache hierarchy set from the architectural vector $\mathbf{a}$ (L1 associativity $2$, L2 associativity $8$).
Workloads are warmed up for $10^{8}$ instructions and measured over the following $5{\times}10^{8}$ instructions per core.
The resulting \texttt{stats.txt} is translated into McPAT's activity schema~\cite{li2009mcpat} configured at the $\SI{45}{\nano\meter}$ node (matching the Cool-3D~\cite{Cool3D} baseline) to emit per-block dynamic, leakage, and gate-leakage power $P_i$ and block areas $A_i$.
This pass produces $\text{IPC}_1(\mathbf{d})$ (the nominal IPC under idealized cache timing, used only as the Level-0 score of~\Cref{subsec:formulation}) and the per-block power and area trace consumed by the thermal lifting.

\paragraph{Lifted output}
The two lifted outputs combine into the realized throughput
\begin{equation}
    \text{BIPS}_2(\mathbf{d};\mathbf{w}) \;=\; \text{IPC}_2(\mathbf{d};\mathbf{w}) \cdot f\!\bigl(T_{\max}(\mathbf{d};\mathbf{w})\bigr),
    \label{eq:bips_method}
\end{equation}
which reproduces~\Cref{eq:bips} of the preliminaries.
In addition to $\text{BIPS}_2$, we retain $\text{IPC}_1$ and $T_{\max}$ so that the Level-0, Level-1, and Level-2 rankings of~\Cref{subsec:formulation} can all be reported over the same design set.
The Kendall-$\tau$ between each pair quantifies how much ranking signal each additional lifting stage contributes; we use this decomposition in~\Cref{sec:exp} as a diagnostic on why the lifting is necessary in the first place.
The analytical floorplanner of~\Cref{sec:analytical-placer} then takes the lifted block representation as its input.

\paragraph{Per-configuration cost}
Each configuration requires two gem5 invocations (dominated by the out-of-order simulation at $\mathcal{O}(I)$ per workload), one McPAT call ($\mathcal{O}(|B|)$), one HotSpot call per layout queried ($\mathcal{O}(G^2)$ for a $G{\times}G$ grid), and an $\mathcal{O}(1)$ CACTI lookup amortized over the design-space exploration (DSE).
gem5 dominates; per-configuration wall time of the lifting stage is $\sim\!\SI{80}{\second}$ on our workstation, and the closed-loop sweep across both cooling envelopes (\Cref{subsec:setup}) completes in about $1\,\mathrm{h}$ per envelope at 16-way parallelism on 32 physical cores.
Per-worker scratch directories isolate McPAT, CACTI, and HotSpot intermediate files so the 24-way speedup runs without cross-worker contention on shared output paths.

\subsection{Thermal Lifting: Stack, Floorplan Materialization, and HotSpot}
\label{subsec:thermal_loop}
Given the per-block power and area trace from the shared front-end, the thermal lifting produces $T_{\max}(\mathbf{d})$ for any layout candidate by rasterizing that layout onto a uniform grid and running HotSpot to steady state.
For diagnostic experiments that fix layout (\Cref{subsec:ranking_baseline,subsec:ranking_stressed}) and as the baseline against which~\Cref{sec:analytical-placer} is compared, we use a deterministic shelf-pack rule.
The analytical floorplanner of~\Cref{sec:analytical-placer} replaces this rule with an L-BFGS-B~\cite{lbfgs_proposed} solver whose objective is the realized BIPS itself.

\paragraph{Shelf-pack baseline floorplan}
The baseline layout is a deterministic shelf-pack: a row-based packing rule from VLSI placement, used as the default convention in architectural-stage 3D-DSE flows that lack a thermal-aware placer. It is fully determined by the core count and die geometry, and is what every comparison in the experimental section uses.
Blocks of each core are packed as a rectangular cluster on the bottom die according to a fixed rule: a $1$-core configuration is centered, $2$ cores are split left and right, and $4$ cores occupy either the four quadrants (on $\SI{12}{\mm}^2$ dies, where quadrant area is sufficient to hold the largest per-core cluster) or two side-by-side strips (on the tighter $\SI{10}{\mm}^2$ dies).
The shared L2 and any other memory-side blocks are placed on the top die, which sits closer to the heat sink in the package stack; this mirrors the physical organization used in commercial 3D-stacked products where the cache tier is stacked on top of the compute tier to exploit the shorter thermal path out of the stack.
The output is the per-block tuple $(x_i,y_i,w_i,h_i,z_i)$ with $z_i\in\{0,1\}$ denoting tier assignment in the two-tier stack.

Die area not covered by any microarchitectural block is filled with dead-silicon tiles carrying zero power.
Leaving those regions empty in the power trace would cause HotSpot to model them as insulating voids and overestimate the peak temperature; the silicon filler restores the correct lateral heat-spreading path without introducing any heat source of its own.

\paragraph{Grid power map}
Following the Cool-3D pipeline, we avoid the common HotSpot ``block count mismatch'' pitfall by rasterizing block powers onto a uniform $G{\times}G$ grid per tier via area-overlap accounting:
\begin{equation}
    \text{grid}_z[\,c_y,c_x\,] \;+=\; \sum_{i\in B_z} \frac{\mathrm{overlap}\bigl(b_i,\,\mathrm{cell}_{c_x,c_y}\bigr)}{A_i}\cdot P_i,
    \label{eq:gridmap}
\end{equation}
where $B_z$ is the block set on tier $z$ and $\mathrm{overlap}(\cdot)$ is the geometric intersection area.
This produces a grid representation with identical shape across configurations with different block counts, allowing direct comparison of $T_{\max}$ across designs.
We use $G{=}32$ in this work, giving $\sim\!\SI{313}{\micro\meter}$ cell pitch on a $\SI{10}{\mm}$ die, fine enough to resolve the fastest thermal gradient observed in our sweep, and at a grid resolution that keeps the per-ptrace line under HotSpot's internal buffer limits.

\paragraph{HotSpot configuration}
HotSpot is invoked in grid mode with \texttt{detailed\_3D} enabled on a five-layer stack (interposer, bottom die, thermal interface material (TIM), top die, top TIM).
We use the default Cool-3D package parameters: $\SI{30}{\mm}$ spreader, $\SI{60}{\mm}$ heat sink, natural-convection ambient of $\SI{45}{\celsius}$, and a silicon thickness of $\SI{50}{\micro\meter}$ per die.
The steady-state solution yields the per-cell temperature grid, from which we define $T_{\max}(\mathbf{d})=\max_{z,c_x,c_y}T_z[c_y,c_x]$.
A design that cannot be cooled below $T_{\text{safe}}$ even at $f_{\min}$ is clipped at $f_{\min}$ rather than removed from the search, so that the comparison across points stays monotone and there is no discontinuity at the feasibility boundary.
The thermal lifting is then complete: the architectural vector $\mathbf{a}$ and the physical vector $\mathbf{p}$ jointly determine $f(T_{\max})$ via the closed-form model derived in~\Cref{subsec:closedform_freq}, which the floorplanner of~\Cref{sec:analytical-placer} optimizes $\mathbf{p}$ against on every candidate layout it considers.

\subsection{Latency Lifting: CACTI-Calibrated Discrete Latency}
\label{subsec:delay_loop}
The latency lifting calibrates IPC against cache and on-chip-network timing derived from device-level simulation, replacing the linear-distance wire-delay proxies used by prior flows.

\paragraph{CACTI-calibrated cache latency}
For each cache geometry in our sweep, we run CACTI~\cite{muralimanohar2007cacti} at the $\SI{45}{\nano\meter}$ node with the same technology parameters used by McPAT in the shared front-end (consistent node, consistent transistor model).
The resulting access time in nanoseconds is converted into an integer gem5 cycle count at the nominal $f_0{=}\SI{2.0}{\giga\hertz}$ clock (see~\Cref{tab:cacti}).
Because cache geometry is part of the architectural vector $\mathbf{a}$, a single precomputed lookup table $\ell_{\text{CACTI}}(s_{L1D},s_{L2})$ covers the entire design space without re-running CACTI during DSE.
\input{tables/cacti_lut}

\paragraph{Discrete topology penalties}
On top of the CACTI access time, we add three integer penalties that correspond to the three dominant sources of on-chip communication delay in a 3D multi-core stack:
\begin{equation}
    \mathbf{L}(\mathbf{d}) \;=\; \underbrace{\ell_{\text{CACTI}}\bigl(s_{L1D},s_{L2}\bigr)}_{\text{cache access}}
    \;+\; \underbrace{(N_{\text{cores}}\!-\!1)}_{\text{L2 arbitration}}
    \;+\; \underbrace{2\cdot h_{\text{TSV}}}_{\text{inter-tier}}
    \;+\; \underbrace{1}_{\text{L1 pipe}},
    \label{eq:latency_model}
\end{equation}
where each term is justified as follows.
\begin{itemize}
    \item $\ell_{\text{CACTI}}$ comes from~\Cref{tab:cacti}. It already absorbs all intra-array wordline and bitline delay so no additional distance scaling is applied.
    \item The $(N_{\text{cores}}{-}1)$ term is a standard busy-waiting bound for an $N$-way shared last-level cache under the common MOESI-like coherence assumption, and matches the worst-case per-request arbitration delay reported by prior L2 miss studies~\cite{wu2005joint_cyw}.
    \item $h_{\text{TSV}}$ is the integer number of TSV crossings on the critical path between the requester and the target L2 bank, derived from the floorplan's tier-assignment vector $\{z_i\}$. Each crossing adds $2$ cycles, which matches the capacitive-load-limited traversal time of a $\SI{50}{\micro\meter}$-pitch TSV in $\SI{45}{\nano\meter}$ silicon reported by~\cite{2021TVLSI-thermal-aware-3dfloorplan-tsv}.
    \item The trailing $1$-cycle term is the L1 pipeline stage, consistent with the gem5 \texttt{SimpleMemory} model.
\end{itemize}
$\mathbf{L}(\mathbf{d})$ is back-annotated into gem5 by overriding the \texttt{tag\_latency}, \texttt{data\_latency}, and \texttt{response\_latency} parameters of each cache node plus the \texttt{xbar latency} between L1 and L2.
The second gem5 pass yields $\text{IPC}_2(\mathbf{d})$ on the same workload trace.

\paragraph{Topology-only scope of $\mathbf{L}(\mathbf{d})$}
$\mathbf{L}(\mathbf{d})$ as defined in~\eqref{eq:latency_model} captures only architectural and topology-level penalties (CACTI cache access, shared-L2 arbitration, integer TSV crossings, L1 pipeline). It is layout-invariant: it does not depend on block $(x,y)$ coordinates or intra-die wire distances.
Three observations justify this decoupling quantitatively:
\begin{enumerate}
    \item \textbf{Scale of in-plane delay.} At $\SI{45}{\nano\meter}$ and $\SI{2}{\giga\hertz}$, RC-limited wire propagation across a $\SI{10}{\milli\meter}$ die is under $\SI{500}{\pico\second}$, that is, one cycle. Compared to an L2 access of $6$ to $10$ cycles from~\Cref{tab:cacti}, any in-plane term is below $15\%$ of the dominant cache latency and well within the empirical IPC sensitivity we bound in~\Cref{sec:exp}.
    \item \textbf{Dominance of TSV capacitance.} In a 3D stack, inter-tier physical distance is below $\SI{100}{\micro\meter}$ and propagation delay is dominated by TSV driver capacitance, not wire length~\cite{2021TVLSI-thermal-aware-3dfloorplan-tsv}. A discrete ``cycles-per-hop'' model captures this far more faithfully than a continuous distance-scaled proxy.
    \item \textbf{NoC hop convention.} In modern multi-core fabrics, horizontal inter-core and core-to-L2 traffic is routed through router-based networks-on-chip whose per-hop delay is fixed by topology rather than by floorplan coordinates~\cite{dally2004principles}. Linear distance is therefore already absent from the latency model of the fabric itself.
\end{enumerate}
In summary, floorplan geometry enters the thermal lifting only, through the grid of~\Cref{eq:gridmap}; latency depends only on architectural and topological quantities through~\Cref{eq:latency_model}.
The two lifted outputs remain methodologically disjoint, and both feed into the realized BIPS through~\Cref{eq:bips_method}.
This disjointness is what lets the floorplanner of~\Cref{sec:analytical-placer} treat $\mathbf{L}(\mathbf{d})$ as a per-architecture constant while it varies $\mathbf{p}$ to optimize $T_{\max}$ without revisiting the latency model.
A complementary layout-dependent in-plane wire-delay term, derived from the same Bakoglu--Meindl model, is introduced separately inside the floorplanner objective of~\Cref{eq:placer_loss} and integer-cycled into gem5 at the R2 back-annotation step of~\Cref{subsec:placer_codesign}; it is not part of $\mathbf{L}(\mathbf{d})$ above.
\subsection{Closed-Form Frequency via Thermal-Operator Linearity}
\label{subsec:closedform_freq}
The piecewise-linear throttle of~\Cref{eq:throttle} is convenient, but its slope $\alpha$ is a fitted constant that carries no direct physical meaning.
We replace it with a closed-form expression by observing that the two physical quantities it compresses, the steady-state response of HotSpot and the power-frequency scaling of McPAT, are both linear.
Composing them produces a temperature field whose dependence on $f$ is equally transparent in closed form.

HotSpot's steady-state solver discretizes the 3D heat equation on the per-tier grid and assembles a sparse thermal-conductance matrix $\mathbf{G}_\theta\in\mathbb{R}^{N\times N}$ whose entries are determined entirely by the floorplan, material stack, TIM, and package.
Solving $\mathbf{G}_\theta\,\mathbf{T} = \mathbf{P} + \mathbf{G}_\theta\,T_{\text{amb}}\mathbf{1}$ is therefore equivalent to applying a fixed linear operator
\begin{equation}
    \mathbf{T} - T_{\text{amb}}\mathbf{1} \;=\; \mathbf{R}_\theta\,\mathbf{P},
    \qquad \mathbf{R}_\theta \triangleq \mathbf{G}_\theta^{-1},
    \label{eq:linear_operator}
\end{equation}
to the per-cell power vector $\mathbf{P}\in\mathbb{R}^{N}$.
The operator $\mathbf{R}_\theta$ does not depend on $\mathbf{P}$ or on $f$; for a given 3D stack it is effectively a constant.
On the McPAT side, per-block power is split into a dynamic component $P_{\text{dyn},b}$ and a leakage component $P_{\text{leak},b}$ (sub-threshold and gate combined).
At fixed supply voltage, synchronous switching scales linearly with the clock frequency~\cite{chandrakasan1992lowpower}, while leakage is $f$-independent.
After rasterization onto the thermal grid via~\Cref{eq:gridmap}, the cell-level power vector therefore admits the exact decomposition
\begin{equation}
    \mathbf{P}(f) \;=\; \mathbf{P}_{\text{leak}} \;+\; \frac{f}{f_0}\,\mathbf{P}_{\text{dyn}},
    \label{eq:power_decomp}
\end{equation}
where $\mathbf{P}_{\text{leak}}$ and $\mathbf{P}_{\text{dyn}}$ are both read off directly from the nominal-frequency output of the shared front-end.

Substituting~\Cref{eq:power_decomp} into~\Cref{eq:linear_operator} and applying superposition gives a temperature field that is affine in $f$:
\begin{equation}
    \mathbf{T}(f) \;=\; T_{\text{amb}}\mathbf{1} \;+\; \mathbf{R}_\theta\,\mathbf{P}_{\text{leak}} \;+\; \frac{f}{f_0}\,\mathbf{R}_\theta\,\mathbf{P}_{\text{dyn}}.
    \label{eq:T_of_f}
\end{equation}
Every cell temperature traces a straight line in $f$ whose intercept is the leakage-only self-heat map and whose slope is the dynamic-only self-heat map evaluated at $f_0$.
A single HotSpot invocation at $f_0$ already yields $\mathbf{T}(f_0)$.
Evaluating~\Cref{eq:T_of_f} at any other frequency is $\mathcal{O}(N)$ vector arithmetic, with no need to re-solve $\mathbf{G}_\theta$ at each trial frequency.

In the absence of per-block leakage detail, a further simplification is both convenient and empirically accurate.
Let
\begin{equation}
    \gamma \;\triangleq\; \frac{\sum_c P_{\text{leak},c}}{\sum_c \bigl(P_{\text{leak},c}+P_{\text{dyn},c}\bigr)}
\end{equation}
denote the grid-level leakage fraction, which McPAT reports per configuration.
If the per-cell ratio $P_{\text{leak},c}/P_c(f_0)$ is approximately uniform across the die, which is reasonable when logic and SRAM share the same process node, then $\mathbf{P}_{\text{leak}}\approx\gamma\mathbf{P}(f_0)$ and $\mathbf{P}_{\text{dyn}}\approx(1-\gamma)\mathbf{P}(f_0)$, and~\Cref{eq:T_of_f} collapses into a rigid scaling of a single map:
\begin{equation}
    \mathbf{T}(f) - T_{\text{amb}}\mathbf{1}
    \;=\;
    \Bigl[\gamma + (1-\gamma)\tfrac{f}{f_0}\Bigr]\bigl(\mathbf{T}(f_0) - T_{\text{amb}}\mathbf{1}\bigr).
    \label{eq:T_scalar}
\end{equation}
The scalar factor is the same at every cell and strictly positive, so the peak location is preserved across $f$ and the peak temperature obeys
\begin{equation}
    T_{\max}(f) - T_{\text{amb}}
    \;=\;
    \Bigl[\gamma + (1-\gamma)\tfrac{f}{f_0}\Bigr]\bigl(T_{\max}(f_0) - T_{\text{amb}}\bigr).
    \label{eq:Tmax_of_f}
\end{equation}

Setting $T_{\max}(f)=T_{\text{safe}}$ in~\Cref{eq:Tmax_of_f} and solving for $f$ gives the highest clock frequency consistent with the thermal budget:
\begin{equation}
    f_{\text{sus}}(\mathbf{d}) \;=\;
    \max\!\left(f_{\min},\;
    \frac{f_0}{1-\gamma}\!\left[\frac{T_{\text{safe}}-T_{\text{amb}}}{T_{\max}(\mathbf{d};\,f_0)-T_{\text{amb}}} - \gamma\right]\right)
    \label{eq:fsus}
\end{equation}
when $T_{\max}(\mathbf{d};f_0)>T_{\text{safe}}$, and $f_{\text{sus}}(\mathbf{d})=f_0$ otherwise.
Equation~\eqref{eq:fsus} replaces the fitted slope $\alpha$ of~\Cref{eq:throttle} with two directly observable parameters: $T_{\text{safe}}$, the dynamic thermal management (DTM) trigger used by commodity DVFS controllers (typically $\SI{85}{\celsius}$ on server-class silicon), and $\gamma$, the leakage fraction read straight out of McPAT.
CLIP-3D uses this form as the definition of $f(T_{\max})$ in the $\text{BIPS}_2$ objective of~\Cref{eq:bips} and in the ranking diagnostics of~\Cref{sec:exp}.

Two consequences follow.
First, because~\Cref{eq:fsus} is closed-form in the single scalar $T_{\max}(\mathbf{d};f_0)$, sweeping $\gamma$ or $T_{\text{safe}}$ at evaluation time costs $\mathcal{O}(|\mathcal{D}|)$ scalar operations rather than $|\mathcal{D}|$ fresh thermal simulations, and the per-configuration HotSpot solve of the thermal lifting is reused.
Second, the uniform-$\gamma$ assumption underlying~\Cref{eq:T_scalar} can be lifted without giving up the closed-form structure: evaluating~\Cref{eq:T_of_f} once with $\mathbf{P}_{\text{dyn}}=\mathbf{0}$ and once with $\mathbf{P}_{\text{leak}}=\mathbf{0}$ calibrates the leakage and dynamic self-heat maps separately, after which the per-cell affine form in $f$ is exact.
This two-point calibration costs one extra HotSpot solve per configuration and serves as the principled fallback when logic and cache have markedly different leakage profiles across block classes.
\input{algorithms/clip3d_placer}

\subsection{Analytical 3D Thermal-Aware Floorplanner with Closed-Form $f_{\text{sus}}$ Inside the Objective}
\label{sec:analytical-placer}

The lifted block representation of \Cref{subsec:overview,subsec:thermal_loop,subsec:delay_loop} and the closed-form sustainable frequency of~\Cref{eq:fsus} together make the inner $\mathbf{p}^{\star}(\mathbf{a})$ of~\Cref{eq:objective} solvable analytically.
We can search over $\mathbf{p}$ for the layout that maximizes the realized BIPS itself, rather than minimizing a surrogate that approximates it.
Prior 3D thermal-aware floorplanning frameworks (e.g.,~\cite{sa3dfp_guan_2023}) optimize a linear cost
$\,w_{\text{HPWL}}\!\cdot\!\text{HPWL}+w_{\text{thermal}}\!\cdot\!T_{\max}\,$
whose weights $(w_{\text{HPWL}},w_{\text{thermal}})$ are not derivable from the architectural quantity the system actually delivers.
The optimum of that linear cost is in general not the optimum of $\text{BIPS}_2 = \mathrm{IPC}\!\cdot\!f_{\text{sus}}(T_{\max}(\mathbf{d}),\gamma)$, and the gap is workload- and regime-dependent (\Cref{sec:exp}).
The linear cost also stays positive on configurations with thermal headroom, where there is no $T_{\max}$ pressure to relieve, so small layout perturbations can move the score in either direction and produce false-positive ``improvements''.
We therefore embed~\Cref{eq:fsus} directly in the floorplanner's objective, so cross-tier macro assignment and in-plane placement are co-optimized for the same realized BIPS the rest of the pipeline scores designs on.

\paragraph{Decision variables.}
For a $K$-tier stack, let $z\!\in\!\{0,1\}^M$ be the cross-tier assignment of $M$ movable macros (the L2 cache and, optionally, large core tiles) and let $(x,y)\!\in\!\mathbb{R}^{2M}$ be their in-plane coordinates.
We treat $z$ discretely via Fiduccia--Mattheyses-style 2-way enumeration over the $K^M$ tier assignments and run the continuous L-BFGS-B optimization conditioned on each candidate.
For the 4-core 2-tier P1 stack of our experiments ($K\!=\!2$, $M\!=\!1$ since the 4 cores are pre-grouped into a tile and only the L2 macro is free to migrate), this enumeration reduces to two L-BFGS-B runs and is exact.
For larger $M$ a standard FM partitioner~\cite{1995ICCAD-FM_partition} replaces the enumeration without changing the rest of the pipeline.

\paragraph{Block-level thermal proxy.}
Within the inner optimization loop we need a temperature signal that is monotone in the right direction (hot blocks clustered together or relegated to the bottom tier produce a higher predicted $T_{\max}$) and differentiable in $(x,y)$, but not a high-fidelity simulator.
We aggregate the per-sub-block $\texttt{ptrace}$ entries into per-tile centroids $(x_t,y_t)$ with summed power $P_t$, and predict
\begin{equation}
\begin{split}
\hat T_{\max}(x_{L2}, y_{L2}; z) \;=\;& T_{\text{amb}} + R_{\text{conv}}P_{\text{tot}} \\
&{}+\, \alpha\,\max_i\!\sum_j P_j\,\mathcal{K}(d_{ij})\,w(z_i,z_j) \\
&{}+\, \beta\!\sum_{j\in\text{bot}}\!P_j,
\end{split}
\label{eq:tmax_proxy}
\end{equation}
where $\mathcal{K}(d)\!=\!1/\sqrt{1+(d/L_c)^2}$ is a smooth Pythagorean kernel with $L_c$ set to the die half-width, $w(z_i,z_j)\!\in\!\{1,w_{\text{cross}}\}$ down-weights cross-tier pairs, $R_{\text{conv}}P_{\text{tot}}$ is the exact lumped-heatsink rise, and $(\alpha,\beta)$ are unit-conversion constants chosen once so $\hat T_{\max}$ lands in the $80$ to $110\,^\circ\text{C}$ regime where~\Cref{eq:fsus} has non-trivial response.
The proxy is intentionally crude: only its gradient direction drives the optimizer, and the absolute $T_{\max}$ that enters all paper-reported numbers comes from the validation HotSpot pass below.
This separation lets the inner loop be HotSpot-free without committing to any particular thermal-model accuracy claim.

\paragraph{Wire-delay term.}
The floorplanner's wire term reuses the Bakoglu--Meindl~\cite{bakoglu1985optimal} distributed-RC line delay $\tau(L)=0.69\,RCL^2$ of~\Cref{subsec:delay_loop}, evaluated at the smoothed Manhattan distance $L_i = \sqrt{(x_{\text{L2}}-x_i)^2+\epsilon^2} + \sqrt{(y_{\text{L2}}-y_i)^2+\epsilon^2} + \ell_{\text{TSV}}$ from each core $i$ to the L2 macro.
The vertical TSV traversal $\ell_{\text{TSV}}$ is constant within a candidate (since L2 always sits on the opposite tier from the cores in our P1 setup) and the smoothing keeps the term differentiable in $(x_{L2}, y_{L2})$.

\paragraph{Objective.}
Combining the two pieces yields the analytical floorplanner's loss, conditioned on a fixed discrete $z$ chosen by the outer FM enumeration:
\begin{equation}
\begin{split}
\mathcal{L}(z; x_{L2}, y_{L2}) \;=\;
&-\underbrace{\mathrm{IPC}\!\cdot\!f_{\text{sus}}\!\bigl(\hat T_{\max}(x_{L2}, y_{L2}; z),\gamma\bigr)}_{\text{BIPS}_2 \text{ proxy (C2 inside obj.)}} \\
&+\, \lambda_{\text{wire}}\!\cdot\!\mathrm{IPC}\!\cdot\!\overline{\tau}(x_{L2}, y_{L2}),
\end{split}
\label{eq:placer_loss}
\end{equation}
where $\overline{\tau}$ is the average $\tau(L_i)$ across cores (in cycles, scaled by $\mathrm{IPC}$ to share the BIPS unit).
We solve~\Cref{eq:placer_loss} with L-BFGS-B subject to the box constraints $(x_{L2},y_{L2})\!\in\![0,W\!-\!w_{\text{L2}}]^2$, multi-started from three initial points (BL corner, die centre, TR corner) for each enumerated $z$.
The global optimum is the best of the $|\{z\}|\!\times\!3$ candidates.
\Cref{alg:clip3d_placer} summarises the full procedure.

\input{tables/closedform_validation_v3p5}

\paragraph{Validation pass.}
The chosen tier assignment $z^\star$ and continuous optimum $(x_{L2}^\star, y_{L2}^\star)$ define a final layout that is materialised by patching the L2-bearing die's \texttt{.flp} and reordering the \texttt{coremem.ptrace} columns.
One real HotSpot run on this layout supplies the true $T_{\max}^\star$ that enters~\Cref{eq:fsus} (\Cref{alg:clip3d_placer}, lines 11--13).
The proxy carries the optimizer through the search; the validation pass guarantees the reported $\text{BIPS}_2^\star$ uses exactly the same thermal model as every other configuration in the design space.

The analytical floorplanner differs from a wrapper around an existing CAD tool in two ways.
First, the closed-form objective is grounded in~\Cref{eq:fsus}: $f_{\text{sus}}$ enters the loss directly, not as a downstream score on a layout that was selected for a different criterion.
Second, no thermal simulator is invoked during the optimization itself.
The inner loop costs only the proxy evaluation and L-BFGS-B's gradient queries, both microsecond-scale, so the wall-time per layout is dominated by the single validation HotSpot rather than by repeated solver calls inside the inner loop.
The $\lambda_{\text{wire}}$ coefficient is the only weight in~\Cref{eq:placer_loss} and serves a specific unit-conversion role (cycles to BIPS) rather than balancing competing surrogate objectives in the DAC sense.
The wire delay itself is parameter-free at the architectural stage, derived from PTM~\cite{zhao2007predictive} 45-nm metal RC ($R{=}50\,\Omega/\text{mm}$, $C{=}200\,\text{fF}/\text{mm}$ for semi-global wires) via Bakoglu--Meindl.

%% file: tables/cacti_lut.tex
\begin{table}[t]
\centering
\caption{CACTI-calibrated cache access latency and area at the $\SI{45}{\nano\meter}$ node, used to populate $\ell_{\text{CACTI}}(s_{L1D},s_{L2})$ in~\Cref{eq:latency_model}.}
\label{tab:cacti}
\footnotesize
\setlength{\tabcolsep}{5pt}
\resizebox{0.75\linewidth}{!}{%
\begin{tabular}{l c c c}
\toprule
\textbf{Cache} & \textbf{Access time} & \textbf{gem5 cycles} & \textbf{Area} \\
\midrule
L1D, 16\,kB   & 0.97 & 2  & 0.52  \\
L1D, 32\,kB   & 1.01 & 3  & 0.74  \\
L1D, 64\,kB   & 1.13 & 3  & 1.16  \\
L1D, 128\,kB  & 1.47 & 3  & 2.25  \\
\midrule
L2, 128\,kB   & 2.66 & 6  & 2.72  \\
L2, 256\,kB   & 2.78 & 6  & 6.29  \\
L2, 512\,kB   & 3.14 & 7  & 10.01 \\
L2, 1\,MB     & 3.96 & 8  & 16.68 \\
L2, 2\,MB     & 4.97 & 10 & 36.99 \\
\bottomrule
\end{tabular}
}

\raggedright
\scriptsize
Access time in \si{\nano\second}; area in $\mathrm{mm}^2$. gem5 cycles are integer-rounded at $f_0{=}\SI{2.0}{\giga\hertz}$.
\vspace{-0.5em}
\end{table}

%% file: algorithms/clip3d_placer.tex
\begin{algorithm}[t!]
    \caption{CLIP-3D floorplanner with closed-form $f_\text{sus}$ inside the objective and a single end-of-search HotSpot validation.}
    \label{alg:clip3d_placer}
    \small
    \setlength{\hsize}{0.95\linewidth}
    \KwIn{base layouts $\mathcal{L}_{z=0}, \mathcal{L}_{z=1}$;\, $\gamma$, $\mathrm{IPC}$, $f_0$, $T_\text{safe}$, $T_\text{amb}$, $f_{\min}$, $\lambda_\text{wire}$}
    \KwOut{optimum tier $z^\star$, in-plane L2 position $(x_{L2}^\star, y_{L2}^\star)$, HotSpot-validated layout $\mathcal{L}^\dagger$}
    $\hat T(x_{L2}, y_{L2}; z) \leftarrow \textsc{ProxThermal}(\mathcal{L}_z)$ \tcp*{\Cref{eq:tmax_proxy}}
    $\hat f_\text{sus}(\hat T) \leftarrow \textsc{ClosedFormFsus}(\hat T, \gamma)$ \tcp*{\Cref{eq:fsus}}
    $\bar\tau(x_{L2}, y_{L2}) \leftarrow \textsc{MeanWireDelay}(x_{L2}, y_{L2}; \mathcal{L}_z)$ \tcp*{Bakoglu--Meindl, \Cref{subsec:delay_loop}}
    $\mathcal{L}(z, x_{L2}, y_{L2}) \leftarrow \textsc{PlacerLoss}(\hat f_\text{sus}, \bar\tau)$ \tcp*{\Cref{eq:placer_loss}}
    $\mathcal{S}^\star \leftarrow \emptyset$\;
    \For{\textup{$z \in \{0, 1\}$}}{
        \For{\textup{each start $(x_0, y_0) \in \{\text{BL}, \text{centre}, \text{TR}\}$}}{
            $\mathbf{u}^\star \leftarrow \textsc{LBFGSB}(\mathcal{L}(z, \cdot, \cdot),\, (x_0, y_0))$ \tcp*{box bounds $[0, d{-}w_{L2}]^2$}
            $\mathcal{S}^\star \leftarrow \mathcal{S}^\star \cup \{(z, \mathbf{u}^\star, \mathcal{L}(z, \mathbf{u}^\star))\}$\;
        }
    }
    $(z^\star, x_{L2}^\star, y_{L2}^\star) \leftarrow \arg\min_{(z, \mathbf{u}, \ell) \in \mathcal{S}^\star} \ell$\;
    $\mathcal{L}^\dagger \leftarrow \textsc{RewriteLayout}(\mathcal{L}_{z^\star}, x_{L2}^\star, y_{L2}^\star)$ \tcp*{materialize layout at the optimum}
    $T_{\max}^\dagger,\, f_\text{sus}^\dagger \leftarrow \textsc{HotSpot}(\mathcal{L}^\dagger)$ \tcp*{single end-of-search validation}
    \Return $(z^\star, x_{L2}^\star, y_{L2}^\star, \mathcal{L}^\dagger, T_{\max}^\dagger, f_\text{sus}^\dagger)$\;
\end{algorithm}

%% file: tables/closedform_validation_v3p5.tex
\begin{table*}[t!]
\centering
\caption{Numerical validation of the closed-form throttle model~\eqref{eq:fsus} across the two cooling envelopes of~\Cref{subsec:setup}.}
\label{tab:closedform_validation_v3p5}
\small
\setlength{\tabcolsep}{5pt}%
\begin{tabular}{l c c c c c c c c c}
\toprule
Workload & $L_2$ & $L_{1D}$ & $\gamma$ & $P(f_0)$ (\si{\watt}) & $T_{\max}(f_0)$ & $f_{\text{sus}}$ (\si{\giga\hertz}) & $T@f_{\text{sus}}$ (\si{\celsius}) & $\max|\Delta_{\text{lin}}|$ & $|\Delta_{\text{safe}}|$\\
\midrule
\multicolumn{10}{l}{\textit{Realistic cooling ($r_{\text{conv}}{=}\SI{3.5}{\kelvin\per\watt}$)}} \\
\cmidrule(lr){1-10}
STREAM   & 256\,kB & 32\,kB & 0.694 & 9.69 & 63.7 & 2.000 & N/A & 0.001 & N/A \\
MATMUL   & 2\,MB & 128\,kB & 0.443 & 16.68 & 93.7 & 2.000 & N/A & 0.013 & N/A \\
FFT      & 256\,kB & 32\,kB & 0.444 & 15.32 & 94.6 & 2.000 & N/A & 0.011 & N/A \\
FFT      & 1\,MB & 64\,kB & 0.446 & 16.24 & 100.0 & 1.760 & 95.00 & 0.012 & 0.000 \\
\midrule
\multicolumn{10}{l}{\textit{Stressed cooling ($r_{\text{conv}}{=}\SI{5.0}{\kelvin\per\watt}$)}} \\
\cmidrule(lr){1-10}
STREAM   & 256\,kB & 32\,kB & 0.694 & 9.69 & 78.2 & 2.000 & N/A & 0.015 & N/A \\
MATMUL   & 2\,MB & 128\,kB & 0.443 & 16.68 & 118.8 & 1.090 & 95.01 & 0.014 & 0.010 \\
FFT      & 1\,MB & 64\,kB & 0.446 & 16.24 & 124.4 & 0.932 & 95.01 & 0.008 & 0.010 \\
\bottomrule
\end{tabular}
\end{table*}

%% file: doc/4-experiment.tex
\section{Experimental Results}
\label{sec:exp}
\input{tables/ranking_combined}
\subsection{Experimental Setup}
\label{subsec:setup}
All experiments run on a Linux workstation with dual Intel Xeon Gold 6426Y CPUs (32 physical cores) and 256\,GB RAM. We adopt the 45\,nm node parameters from Cool-3D~\cite{Cool3D} and match McPAT, CACTI, and HotSpot to the same node for consistency across the pipeline.
\paragraph{Architectural sweep}
We evaluate five workloads on a uniform architectural grid: FFT and CHOLESKY from SPLASH-2~\cite{splash2_1995} (mixed compute with irregular memory, and sparse Cholesky factorization on the \texttt{tk14} input, respectively), the McCalpin STREAM~\cite{mccalpin1995stream} memory-bandwidth benchmark, dense matrix multiplication (MATMUL), and a 2D Jacobi five-point stencil (STENCIL). The mix covers compute, bandwidth, sparse, and stencil regimes.
The architectural sweep varies L2 size $\in\{128,\,256,\,512,\,1024,\,2048\}\,\text{kB}$ (CACTI access time spans 6 to 10 cycles at $f_0$) and L1D size $\in\{16,\,32,\,64,\,128\}\,\text{kB}$, giving $5\times4=20$ architectural configurations per workload and $5\times20=100$ closed-loop points per cooling envelope. Core count is fixed at four for this study; the 2-tier stack uses pattern P1 (cores on the bottom die, L2 on the top die) with auto die sizing at 70\,\% utilization. McPAT-derived per-block area is rescaled by a single global calibration factor so that the reference design (4 cores, 32\,kB L1D, 512\,kB L2 at 45\,nm) lands at 150\,$\text{mm}^2$, matching Intel Nehalem-EP minus its L3 region.
\paragraph{Cooling envelopes}
We evaluate two cooling envelopes with the same package and silicon limits ($T_{\text{safe}}{=}\SI{95}{\celsius}$ silicon $T_j^{\max}$, $T_{\text{amb}}{=}\SI{25}{\celsius}$ data-center inlet, $f_0{=}\SI{2.0}{\giga\hertz}$, $f_{\min}{=}\SI{0.4}{\giga\hertz}$, per-config $\gamma$ from McPAT runtime power):
\begin{itemize}
\item \textbf{Baseline}, $r_{\text{conv}}{=}\SI{3.5}{\kelvin\per\watt}$. Tower heatsink on the package, the nominal operating envelope.
\item \textbf{Stressed}, $r_{\text{conv}}{=}\SI{5.0}{\kelvin\per\watt}$. Reduced cooling headroom, modeling a degraded heatsink such as an aged or partly clogged tower fan, or operation closer to the package thermal limit. This value matches the floorplanner study of~\Cref{subsec:placer_codesign}.
\end{itemize}
Both envelopes use the same $T_{\text{safe}}$, ambient, and architectural sweep. The two-envelope design tests whether IPC ranks remain stable when sustained frequency moves under tighter heat removal.

\paragraph{Per-config pipeline}
Each closed-loop point invokes two gem5 passes (front-end for $\text{IPC}_1$ and back-annotated rerun for $\text{IPC}_2$ under CACTI-tuned cache latency), one McPAT power pass, one HotSpot steady-state solve at $f_0$, and one closed-form $f_{\text{sus}}$ evaluation per~\Cref{eq:fsus}. Total simulation volume across the two envelopes is $200$ completed closed-loop points; the sweep completes in about $1$\,h at 16-way parallelism using per-worker scratch directories to isolate transient files.

\subsection{Closed-Form Frequency Validation}
\label{subsec:closedform_val}
Before reporting rankings built on the closed-form $f_{\text{sus}}$ of~\Cref{eq:fsus}, we confirm that the linearity argument of~\Cref{subsec:closedform_freq} (temperature affine in $f$ under the uniform-$\gamma$ assumption~\eqref{eq:T_scalar}) holds numerically under both cooling envelopes.

We pick four anchor configurations spanning the full thermal regime under the baseline envelope: a cool STREAM point ($T_{\max}{=}\SI{63.7}{\celsius}$, $\gamma{=}0.694$), a borderline-no-throttle MATMUL ($\SI{93.7}{\celsius}$, $\gamma{=}0.443$), an FFT at the throttle boundary ($\SI{94.6}{\celsius}$, $\gamma{=}0.444$), and a heavily throttled FFT ($\SI{100.0}{\celsius}$, $\gamma{=}0.446$). At each anchor we re-run HotSpot at $f \in \{0.5,1.0,2.0\}\,\si{\giga\hertz}$, scaling the ptrace by $\gamma{+}(1{-}\gamma)(f/f_0)$. We additionally rerun the STREAM, MATMUL, and heavily throttled FFT anchors under the stressed envelope, where the MATMUL anchor crosses from headroom into throttling and the FFT anchor enters deeper throttling. \Cref{tab:closedform_validation_v3p5} reports both envelopes, with the per-anchor linearity error $\Delta_{\text{lin}} = T_{\max}^{\text{HotSpot}}(f) - (T_{\text{amb}} + R_\theta P(f))$ taken across $f \in \{0.5,1.0,2.0\}$\,\si{\giga\hertz}, and the post-solve safety margin $\Delta_{\text{safe}} = |T_{\max}^{\text{HotSpot}}(f_{\text{sus}}) - T_{\text{safe}}|$. Rows with $f_{\text{sus}}{=}f_0{=}\SI{2.0}{\giga\hertz}$ have $T_{\max}(f_0){<}T_{\text{safe}}$ at the nominal frequency, so no throttling applies and $T@f_{\text{sus}}$ and $|\Delta_{\text{safe}}|$ are reported as N/A.

The linearity prediction $T_{\max}(f){=}T_{\text{amb}}+R_\theta P(f)$ matches the measured HotSpot output to within $\SI{0.015}{\celsius}$ across all $(4{+}3)\!\times\!3{=}21$ runs, at the level of HotSpot's Gauss--Seidel convergence tolerance. Under the baseline envelope the heavily throttled FFT anchor solves $f_{\text{sus}}{=}\SI{1.760}{\giga\hertz}$ and HotSpot drives $T_{\max}$ to exactly $T_{\text{safe}}{=}\SI{95.00}{\celsius}$, with $|\Delta_{\text{safe}}|{=}\SI{0.0}{\celsius}$. Under the stressed envelope the same FFT anchor solves $f_{\text{sus}}{=}\SI{0.93}{\giga\hertz}$ and the regime-transition MATMUL anchor solves $f_{\text{sus}}{=}\SI{1.09}{\giga\hertz}$, both landing within $\SI{0.01}{\celsius}$ of $T_{\text{safe}}$. The closed-form claim of~\Cref{subsec:closedform_freq} therefore extends to both cooling regimes without modification, and we use it directly in the ranking experiments of the next two subsections.

\subsection{Ranking Diagnostic Across Two Cooling Envelopes}
\label{subsec:ranking_baseline}

We first quantify how much ranking information an IPC-only evaluator discards under baseline cooling, then repeat the same diagnostic under the stressed envelope to test how the picture changes with cooling headroom. For each of the 100 design points (five workloads at the full 20-config grid each) we record $\text{IPC}_1$, $T_{\max}(f_0)$, the closed-form $f_{\text{sus}}$ from~\Cref{eq:fsus}, and the closed-loop $\text{BIPS}_2{=}\text{IPC}_2\!\cdot\!f_{\text{sus}}$. We compare three rankings: $\pi_0$ from $\text{IPC}_1$, $\pi_1$ from $\text{IPC}_1\!\cdot\!f_{\text{sus}}$, and $\pi_2$ from $\text{BIPS}_2$. \Cref{tab:ranking_v3p5} reports, per workload, the number of configs $n$; the throttled fraction (configs with $T_{\max}{>}T_{\text{safe}}$, equivalently $f_{\text{sus}}{<}f_0$); the $\text{BIPS}_2$-rank of the IPC-best design (column IPC\#1$\rightarrow$BIPS$_2$, ideal value $1$, range $[1,n]$); the top-10 overlap $|\text{Top}_{10}(\pi_0)\cap\text{Top}_{10}(\pi_2)|$; the Kendall-$\tau$ pair $\tau_{0,2}$ between $\pi_0$ and $\pi_2$ and $\tau_{1,2}$ between $\pi_1$ and $\pi_2$ (the two coincide in this study because no per-method R2 separates them); and the best-config $\text{BIPS}_2$ ratio of the BIPS-best over the IPC-best design (column ``Best gap'', geometric mean across workloads in summary rows).

\input{tables/tab_full_r2_main}
\input{tables/tab_method_summary}
\input{tables/tab_paper_vs_real}

Under baseline cooling, \Cref{tab:ranking_v3p5} reports the per-workload mismatch. Two regimes appear. CHOLESKY, MATMUL, STENCIL, and STREAM retain full thermal headroom across all 20 configs ($f_{\text{sus}}{=}f_0$ throughout), so $\pi_0$, $\pi_1$, and $\pi_2$ collapse to the same ranking up to a frequency constant. FFT is the only workload with throttled configs ($75\,\%$), and accordingly is the one workload where the IPC-best design is not the $\text{BIPS}_2$-best: it lands at $\tau_{0,2}{=}+0.095$ with the IPC-best design at $\#11$ in $\text{BIPS}_2$. This baseline result is useful as a control: the corrections do not bind on workloads that retain headroom, but they already bind on FFT, whose power map exceeds the package envelope on the larger-L2 configurations at nominal frequency.

\label{subsec:ranking_stressed}
We then re-run the same architectural sweep, layout model, power model, and temperature limit under the stressed envelope ($r_{\text{conv}}{=}\SI{5.0}{\kelvin\per\watt}$); the only quantity that changes is the package thermal resistance. Under stressed cooling, the throttle rate rises across all five workloads, and the ranking metrics that were near-trivial under the baseline envelope now diverge.

\Cref{tab:ranking_stressed_v4} reports the per-workload metrics in the lower block of the table. The throttle rate climbs from a baseline-envelope mean of $15\,\%$ to $68\,\%$ under stressed cooling, with two workloads (CHOLESKY, FFT) at $100\,\%$ and only STREAM remaining mostly in the headroom regime ($20\,\%$). On the throttled side the ranking metrics now diverge: CHOLESKY collapses to $\tau_{0,2}{=}-0.358$ with the IPC-best design ranking $\#17$ in $\text{BIPS}_2$ and a top-10 overlap of $2/10$; FFT lands at $\tau_{0,2}{=}+0.105$ with the IPC-best design at $\#16$. CHOLESKY has $\tau_{0,2}{<}0$ outright, meaning that picking by IPC$_1$ ranks designs strictly worse than a uniform random permutation. Across the five workloads the geometric-mean best-config $\text{BIPS}_2$ gap of IPC-only selection is $1.07\times$ with the largest individual workload at $1.12\times$ (MATMUL).

The qualitative picture is consistent across the two envelopes. When sustained frequency does not move (baseline), IPC is a serviceable proxy for BIPS. When sustained frequency moves with the architectural choice (stressed), IPC and BIPS diverge, and the divergence carries the full weight of the post-correction structure $\text{BIPS}_2{=}\text{IPC}_2\!\cdot\!f_{\text{sus}}$, where both factors are configuration-dependent. The stressed envelope is therefore best read as a reduced-headroom test rather than a more realistic baseline, and the result supports two interpretations. In any deployment that runs near the package thermal limit (degraded cooling, high ambient, sustained AVX or vector loads), the IPC-only ranking is no longer aligned with what silicon delivers. Even in deployments that operate well under $T_{\text{safe}}$ on average, the BIPS-best architecture under the worst-case cooling differs from the IPC-best, and a designer who picks at the IPC stage forecloses on options that would have won under the binding constraint. Either reading points to the same engineering implication: the two corrections need to be applied before the architecture commitment, not after.

\subsection{Layout Co-Design Using the Closed-Form Frequency}
\label{subsec:placer_codesign}
The architectural sweep above evaluates each design with a fixed shelf-pack layout. The layout choice is itself a degree of freedom, and the same closed-form $f_{\text{sus}}$ that drives the ranking experiments above can serve as the thermal term inside an analytical floorplanner objective. We run four floorplanners on the stressed-cooling sweep over the four workloads (FFT, MATMUL, STENCIL, STREAM) that complete the gem5 R2 back-annotation pipeline at the full L2$\times$L1D grid, $4\times20=80$ configurations in total: the deterministic shelf-pack baseline (cores on the bottom tier, L2 at the bottom-left corner), the cool3d-canonical layout, an SA-plus-$\lambda$-mix grid search that reproduces the prior 3D-IC thermal-aware floorplanning template, and the CLIP-3D analytical floorplanner of~\Cref{alg:clip3d_placer}. The ranking diagnostic of~\Cref{subsec:ranking_stressed} additionally covers CHOLESKY using $\text{IPC}_1$, which is layout-invariant; the layout-side experiment in this subsection is restricted to the four R2-complete workloads. Each method emits a layout, then a single HotSpot pass yields the realized $T_{\max}$ and the closed-form throttled clock $f_{\text{sus}}$ from~\Cref{eq:fsus}.

The closed-loop BIPS depends on both the throttled clock and the IPC delivered under the layout's wire and TSV penalty. The ranking diagnostic above uses $\text{BIPS}_2{=}\text{IPC}_1\!\cdot\!f_{\text{sus}}$ with $\text{IPC}_1$ held constant across methods, since gem5 R1 runs before any layout choice. To verify that the layout-conditional IPC also moves in the same direction we re-run gem5 with each method's optimized wire and TSV cycle counts back-annotated into the cache latency. The realized score is then $\text{BIPS}_{\text{real}}{=}\text{IPC}_{\text{R2}}\!\cdot\!f_{\text{sus}}$.

\Cref{tab:full_r2_main} summarizes the realized BIPS gap over shelf-pack across the 80-configuration sweep, broken down by workload and regime. CLIP-3D wins on every thermally-binding configuration with mean $\Delta\text{BIPS}_{\text{real}}{=}{+}13.75\,\%$ and max ${+}24.77\,\%$ (FFT at $L_2{=}128$\,kB). On the headroom subset, where the shelf-pack layout does not throttle and the closed-form $f_{\text{sus}}{=}f_0$ collapses the thermal term, the wire term in the floorplanner objective still pulls L2 closer to the cores. Realized IPC therefore rises and CLIP-3D delivers a mean $\Delta\text{BIPS}_{\text{real}}{=}{+}5.82\,\%$ even in the headroom regime, with a maximum of $79.0\,\%$ on a STREAM configuration whose shelf-pack placed L2 at the worst-case wire distance. The conservative $\text{IPC}_1$ scoring used in the ranking diagnostic undersold this gain because $\text{IPC}_1$ is layout-invariant by construction.

\Cref{tab:method_summary} compares all three thermal-aware floorplanners on the same sweep against shelf-pack. Cool3d-canonical and SA + $\lambda$-mix both reach essentially the same realized BIPS plateau (${+}10.59\,\%$ and ${+}10.01\,\%$ on average), but pay $2.0\times$ and $2.9\times$ more wall time per configuration than CLIP-3D. The reason is that those methods evaluate three HotSpot candidates each (top-$K$ winners of an 11-point $\lambda$-grid surrogate search), while CLIP-3D pays one HotSpot validation. The closed-form $f_{\text{sus}}$ inside the CLIP-3D objective predicts $T_{\max}$ at the optimum well enough that a single confirmation suffices. CLIP-3D therefore sits on the Pareto frontier of the (wall, realized BIPS) plane: the same wall as the deterministic shelf-pack and the same realized BIPS as the surrogate methods.

\Cref{tab:paper_vs_real} quantifies the gap between the conservative paper formula and the realized BIPS for CLIP-3D. The paper formula reports ${+}2.52\,\%$ mean because $\text{IPC}_1$ is held fixed across methods. Once each method's wire and TSV penalty is back-annotated into the cache latency, the realized gap widens to ${+}10.58\,\%$. The mismatch is sharpest in the headroom bucket: the paper formula scores $0/32$ wins because $\text{IPC}_1$ is layout-invariant and $f_{\text{sus}}{=}f_0$ for every non-throttled config, so paper-BIPS is identical across methods, while R2 back-annotation exposes the wire-driven gain as $26/32$. The paper formula is therefore a conservative pre-verification estimate, while the realized number is the simulator-backed estimate after layout-dependent cache and wire latency are applied.

The realized gap is wider than the paper formula predicts for a concrete reason. $\text{IPC}_1$ uses gem5's default cache latency, fixed at the architecture stage and independent of layout, while $\text{IPC}_{\text{R2}}$ uses the CACTI base latency plus the integer wire and TSV cycles derived from the chosen layout, so a tighter layout (shorter average core-to-L2 wire) reduces effective L2 hit time and lifts IPC. The shelf-pack baseline places L2 in the bottom-left corner, the worst-case wire-distance configuration, so any layout that moves L2 closer to the core cluster pays this penalty back. CLIP-3D's L-BFGS-B inner loop optimizes the average wire term in addition to the closed-form $f_{\text{sus}}$ term, which is why it picks up wire-driven IPC even when the configuration is not thermally binding.
These two mechanisms ($f_{\text{sus}}$ gain and wire optimization) split differently across workloads, and~\Cref{tab:full_r2_main} resolves them per workload. Throttled rows are dominated by the $f_{\text{sus}}$ gain, with FFT the largest at mean $19.7\,\%$ and max $24.8\,\%$. Headroom rows are smaller and dominated by the wire-term contribution; STREAM at ${+}7.4\,\%$ headroom mean comes entirely from wire optimization since $f_{\text{sus}}{=}f_0$ collapses the thermal term, and the $79.0\,\%$ outlier comes from one configuration whose shelf-pack placed L2 at the worst-case wire distance.

Taken together, the ranking diagnostic of~\Cref{subsec:ranking_baseline} and the layout co-design above answer two complementary questions on the same physics-aware pipeline. The ranking diagnostic shows that IPC-only selection fails when architectural choices change both instruction throughput and sustainable frequency, with the failure growing as cooling headroom shrinks. The layout co-design shows that the same closed-form $f_{\text{sus}}$ that diagnoses this failure also drives a single-shot floorplanner that matches surrogate-search BIPS at one-third the wall time. The corrections are not bolted on after the fact; they are the architecture-to-physical lifting itself.

%% file: tables/ranking_combined.tex
\begin{table*}[t]
\centering
\caption{Ranking mismatch on the L2$\times$L1D sweep per workload, under both cooling envelopes of~\Cref{subsec:setup}.}
\label{tab:ranking_v3p5}\label{tab:ranking_stressed_v4}
\small
\setlength{\tabcolsep}{6pt}%
\begin{tabular}{l c c c c c c c c}
\toprule
\textbf{Workload} & $\boldsymbol{n}$ & \textbf{Throttled} & $\boldsymbol{f_{\text{sus}}^{\min}}$ \textbf{(GHz)} & \textbf{IPC\#1$\rightarrow$BIPS$_2$} & \textbf{Top-10} & \textbf{$\tau_{0,2}$} & \textbf{$\tau_{1,2}$} & \textbf{Best gap} \\
\midrule
\multicolumn{9}{l}{\textit{Baseline cooling ($r_{\text{conv}}{=}\SI{3.5}{\kelvin\per\watt}$)}} \\
\cmidrule(lr){1-9}
CHOLESKY & 20 & 0\% & 2.000 & \#2 & 9\,/\,10 & $+0.821$ & $+0.821$ & 1.03$\times$ \\
FFT      & 20 & 75\% & 1.760 & \#11 & 5\,/\,10 & $+0.095$ & $+0.611$ & 1.17$\times$ \\
MATMUL   & 20 & 0\% & 2.000 & \#3 & 7\,/\,10 & $+0.758$ & $+0.758$ & 1.01$\times$ \\
STENCIL  & 20 & 0\% & 2.000 & \#2 & 10\,/\,10 & $+0.716$ & $+0.716$ & 1.09$\times$ \\
STREAM   & 20 & 0\% & 2.000 & \#2 & 8\,/\,10 & $+0.537$ & $+0.537$ & 1.08$\times$ \\
\textbf{Mean} & \textbf{100} & \textbf{15\%} & \textbf{1.760} & \textbf{\#4.0} & \textbf{8\,/\,10} & $\mathbf{+0.585}$ & $\mathbf{+0.688}$ & \textbf{1.07$\times$} \\
\midrule
\multicolumn{9}{l}{\textit{Stressed cooling ($r_{\text{conv}}{=}\SI{5.0}{\kelvin\per\watt}$)}} \\
\cmidrule(lr){1-9}
CHOLESKY & 20 & 100\% & 1.181 & \#17 & 2\,/\,10 & $-0.358$ & $+1.000$ & 1.11$\times$ \\
FFT      & 20 & 100\% & 0.932 & \#16 & 4\,/\,10 & $+0.105$ & $+1.000$ & 1.08$\times$ \\
MATMUL   & 20 & 80\% & 1.090 & \#3 & 5\,/\,10 & $+0.326$ & $+1.000$ & 1.12$\times$ \\
STENCIL  & 20 & 40\% & 1.143 & \#8 & 10\,/\,10 & $+0.853$ & $+1.000$ & 1.04$\times$ \\
STREAM   & 20 & 20\% & 1.801 & \#3 & 10\,/\,10 & $+0.947$ & $+1.000$ & 1.02$\times$ \\
\textbf{Mean} & \textbf{100} & \textbf{68\%} & \textbf{0.932} & \textbf{\#9.4} & \textbf{6\,/\,10} & $\mathbf{+0.375}$ & $\mathbf{+1.000}$ & \textbf{1.07$\times$} \\
\bottomrule
\end{tabular}
\vspace{-0.5em}
\end{table*}

%% file: tables/tab_full_r2_main.tex
\begin{table*}[t]
\centering
\caption{Realized BIPS gain over the shelf-pack baseline on the 80-configuration R2 sweep, broken down by workload and regime (throttled vs.\ headroom). The shelf-pack column reports absolute BIPS$_{\text{real}}$ (mean\,/\,max in GIPS) as the baseline anchor.}
\label{tab:full_r2_main}
\small
\setlength{\tabcolsep}{4pt}
\begin{tabular}{l c c c c c c}
\toprule
& & & \textbf{Baseline} & \multicolumn{3}{c}{\textbf{Method (mean / max $\Delta$\,\% real BIPS, wins)}} \\
\cmidrule(lr){4-4} \cmidrule(lr){5-7}
\textbf{Workload} & \textbf{Regime} & \textbf{n} & \textbf{shelf-pack \textit{(GIPS)}} & \textbf{cool3d-canonical} & \textbf{SA + $\lambda$-mix} & \textbf{CLIP-3D (ours)} \\
\midrule
  FFT & throttled & 20 & 11.36\,/\,13.10 & +19.61\,/\,+24.77\,(20/20) & +18.99\,/\,+24.80\,(20/20) & \textbf{+19.71\,/\,+24.77\,(20/20)} \\
\addlinespace[2pt]
  MATMUL & throttled & 16 & 6.60\,/\,8.26 & +7.73\,/\,+17.98\,(16/16) & +6.91\,/\,+14.72\,(16/16) & \textbf{+7.71\,/\,+14.62\,(16/16)} \\
  MATMUL & headroom & 4 & 3.58\,/\,4.13 & +12.07\,/\,+31.88\,(4/4) & +12.07\,/\,+31.88\,(4/4) & \textbf{+12.07\,/\,+31.88\,(4/4)} \\
\addlinespace[2pt]
  STENCIL & throttled & 8 & 11.40\,/\,12.29 & +16.03\,/\,+17.75\,(8/8) & +13.73\,/\,+15.03\,(8/8) & \textbf{+15.53\,/\,+17.88\,(8/8)} \\
  STENCIL & headroom & 12 & 5.61\,/\,5.95 & +1.65\,/\,+2.68\,(12/12) & +1.65\,/\,+2.68\,(12/12) & \textbf{+1.65\,/\,+2.68\,(12/12)} \\
\addlinespace[2pt]
  STREAM & throttled & 4 & 12.37\,/\,13.30 & +4.19\,/\,+4.51\,(4/4) & +3.63\,/\,+3.91\,(4/4) & \textbf{+4.50\,/\,+4.70\,(4/4)} \\
  STREAM & headroom & 16 & 3.49\,/\,6.14 & +7.39\,/\,+79.02\,(10/16) & +7.39\,/\,+79.02\,(10/16) & \textbf{+7.39\,/\,+79.02\,(10/16)} \\
\addlinespace[2pt]
\midrule
  \textbf{Overall} & throttled & \textbf{48} & 9.86\,/\,13.30 & +13.77\,/\,+24.77\,(48/48) & +12.81\,/\,+24.80\,(48/48) & \textbf{+13.75\,/\,+24.77\,(48/48)} \\
  \textbf{Overall} & headroom & \textbf{32} & 4.29\,/\,6.14 & +5.82\,/\,+79.02\,(26/32) & +5.82\,/\,+79.02\,(26/32) & \textbf{+5.82\,/\,+79.02\,(26/32)} \\
\bottomrule
\end{tabular}
\end{table*}

%% file: tables/tab_method_summary.tex
\begin{table}[t]
\centering
\caption{Per-method summary on the 80-configuration sweep against shelf-pack.}
\label{tab:method_summary}
\small
\setlength{\tabcolsep}{3pt}
\resizebox{\linewidth}{!}{%
\begin{tabular}{l r r r r r}
\toprule
\textbf{Method} & \textbf{real mean} & \textbf{real max} & \textbf{paper mean} & \textbf{paper max} & \textbf{wall (s)} \\
\midrule
  cool3d-canonical & +10.59\,\% & +79.02\,\% & +2.42\,\% & +7.46\,\% & 443.8 \\
  SA + $\lambda$-mix & +10.01\,\% & +79.02\,\% & +2.62\,\% & +7.86\,\% & 642.7 \\
  CLIP-3D (ours) & \textbf{+10.58\,\%} & \textbf{+79.02\,\%} & \textbf{+2.52\,\%} & \textbf{+7.71\,\%} & \textbf{220.1} \\
\bottomrule
\end{tabular}%
}
\end{table}

%% file: tables/tab_paper_vs_real.tex
\begin{table}[t]
\centering
\caption{Conservative paper formula ($\text{IPC}_1\!\cdot\!f_\text{sus}$) vs.\ realized BIPS (gem5 R2) for CLIP-3D over shelf-pack.}
\label{tab:paper_vs_real}
\small
\begin{tabular}{l r r}
\toprule
\textbf{CLIP-3D vs.\ shelf} & \textbf{Paper} & \textbf{Real (R2)} \\
\midrule
  Mean $\Delta\,\%$ overall & +2.52\,\% & \textbf{+10.58\,\%} \\
  Max $\Delta\,\%$ overall & +7.71\,\% & \textbf{+79.02\,\%} \\
  Throttled wins & 48/48 & \textbf{48/48} \\
  Headroom wins  & 0/32 & \textbf{26/32} \\
\bottomrule
\end{tabular}
\end{table}

%% file: doc/5-conclusion.tex
\section{Conclusion}
\label{sec:conclusion}

We presented CLIP-3D, an architecture-to-physical lifting pipeline coupled with an analytical 3D thermal-aware floorplanner whose objective embeds a closed-form sustained-frequency expression in place of the HPWL-plus-temperature linear surrogate that prior 3D thermal-aware floorplanners optimize. Across an 80-configuration sweep under stressed cooling, the floorplanner lies on the wall-time vs.\ realized $\text{BIPS}_2$ Pareto frontier against shelf-pack, cool3d-canonical, and an SA $+\,\lambda$-mix HPWL/$T_{\max}$ surrogate baseline analogous to recent SA-based 3D thermal-aware floorplanners.

%% file: doc/6-acknowledgments.tex
\section*{Acknowledgment}
This work was conducted in the JC STEM Lab of Intelligent Design Automation, funded by The Hong Kong Jockey Club Charities Trust.
This work was supported in part by the Research Grants Council of the Hong Kong Special Administrative Region, China, under Grant Nos.\ CUHK14211324 and CUHK7010840, and in part by ACCESS -- AI Chip Center for Emerging Smart Systems, supported by the InnoHK initiative of the Innovation and Technology Commission of the Hong Kong Special Administrative Region Government.
AI assistants were used for language polishing and partial code generation under the authors' supervision. All research decisions and technical contributions were made by the authors.

%% file: doc/bio.tex
\vspace{-.2in}
\begin{IEEEbiography}
    [{\includegraphics[width=1.0in,height=1.26in,clip,keepaspectratio]{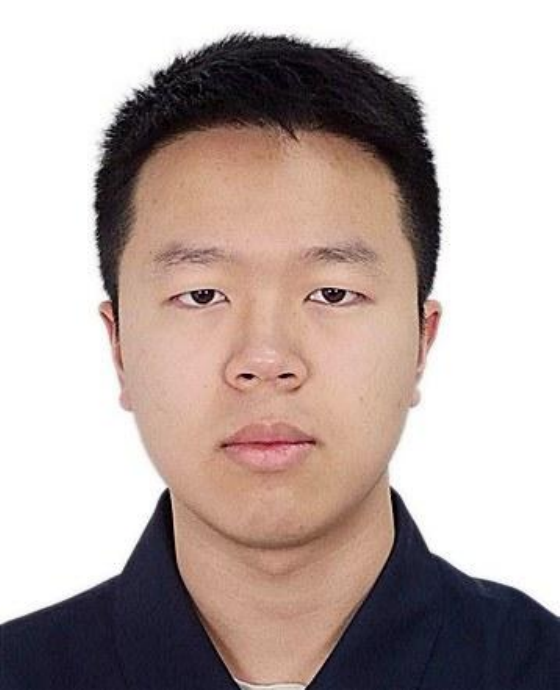}}]
    {Shuo~Ren}
    received his B.S. degree in computer science from the Huazhong University of Science and Technology, Wuhan, China, in 2024. He is currently working toward the Ph.D. degree in the Department of Computer Science and Engineering, The Chinese University of Hong Kong. His current research focuses on electronic design automation, with particular interests in 3D integrated circuit design and physical design optimization.
\end{IEEEbiography}
\vspace{-.5in}
\begin{IEEEbiography}
    [{\includegraphics[width=1.0in,height=1.26in,clip,keepaspectratio]{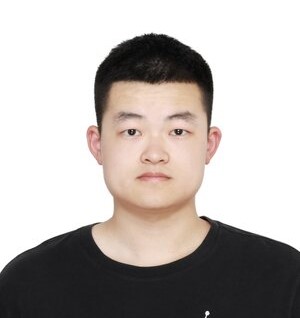}}]
    {Libo~Shen}
    received his B.S. degree in communication engineering from Beijing University of Posts and Telecommunications, Beijing, China, in 2021 and his M.S. degree in computer technology from the University of Chinese Academy of Sciences, Beijing, China, in 2024. He is currently a Ph.D. student at the Department of Computer Science and Engineering, The Chinese University of Hong Kong. His research interests include electronic design automation and computer architecture.
\end{IEEEbiography}
\vspace{-.4in}
\begin{IEEEbiography}
    [{\includegraphics[width=1.0in,height=1.26in,clip,keepaspectratio]{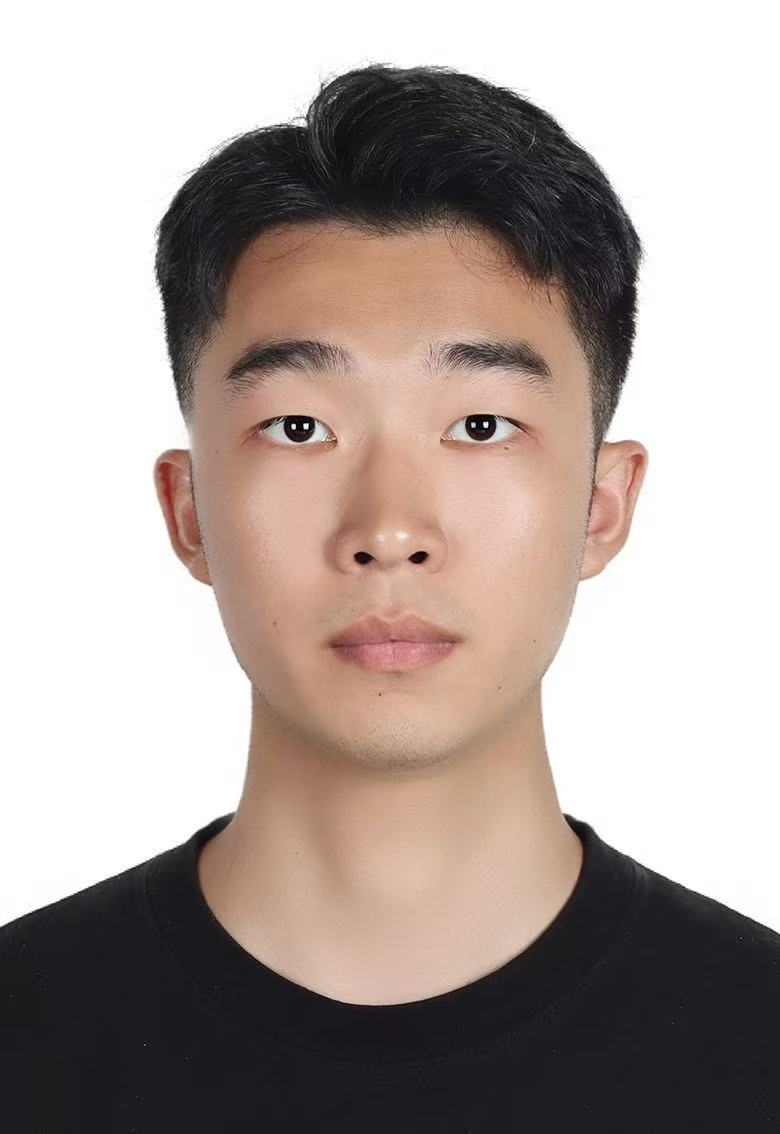}}]
    {Yaohui~Han}
    is currently pursuing the Ph.D. degree with the Department of Computer Science and Engineering, The Chinese University of Hong Kong, under the supervision of Prof.~Tsung-Yi Ho. He received the B.Sc. degree in physics from Central South University, Changsha, China. His current research interest is electronic design automation (EDA), especially EDA for physical design, and large language model-assisted EDA.
\end{IEEEbiography}
\vspace{-.3in}
\begin{IEEEbiography}
    [{\includegraphics[width=1.0in,height=1.26in,clip,keepaspectratio]{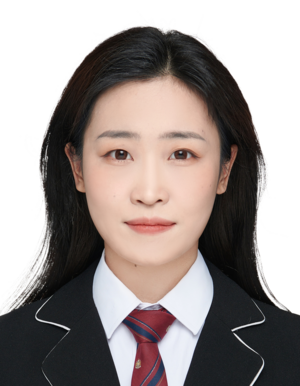}}]
    {Leilei~Jin}
    is currently a postdoctoral researcher affiliated with the Department of Computer Science and Engineering at The Chinese University of Hong Kong. She received her Ph.D. degree from Southeast University in 2024, where her doctoral work focused on foundational theories and practical methodologies in integrated circuit design automation. Her research interests include static timing analysis, crosstalk prediction, PPA optimization for Backside PDN and 3DICs under advanced process nodes.
\end{IEEEbiography}
\vspace{-.2in}
\begin{IEEEbiography}
    [{\includegraphics[width=1.0in,height=1.26in,clip,keepaspectratio]{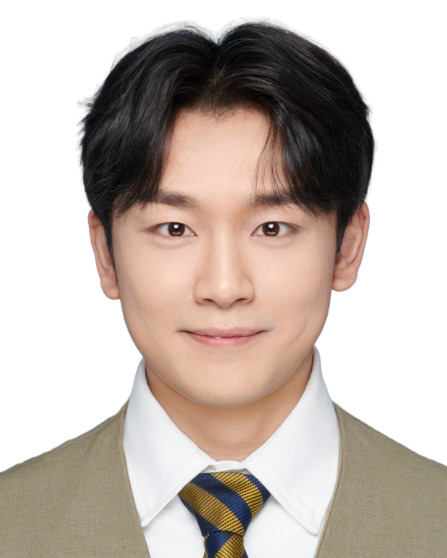}}]
    {Chenghan~Wang}
    received the B.E. degree in electronic science and technology from Harbin Institute of Technology (Weihai) in 2021, and the M.E. degree in circuits and systems from the Institute of Microelectronics, Chinese Academy of Sciences in 2024. He is currently working toward the Ph.D. degree at The Chinese University of Hong Kong. His research interests include electronic design automation (EDA), especially multiphysics simulation, numerical computing, and GPU acceleration.
\end{IEEEbiography}
\vspace{-.2in}
\begin{IEEEbiography}
    [{\includegraphics[width=1.0in,height=1.26in,clip,keepaspectratio]{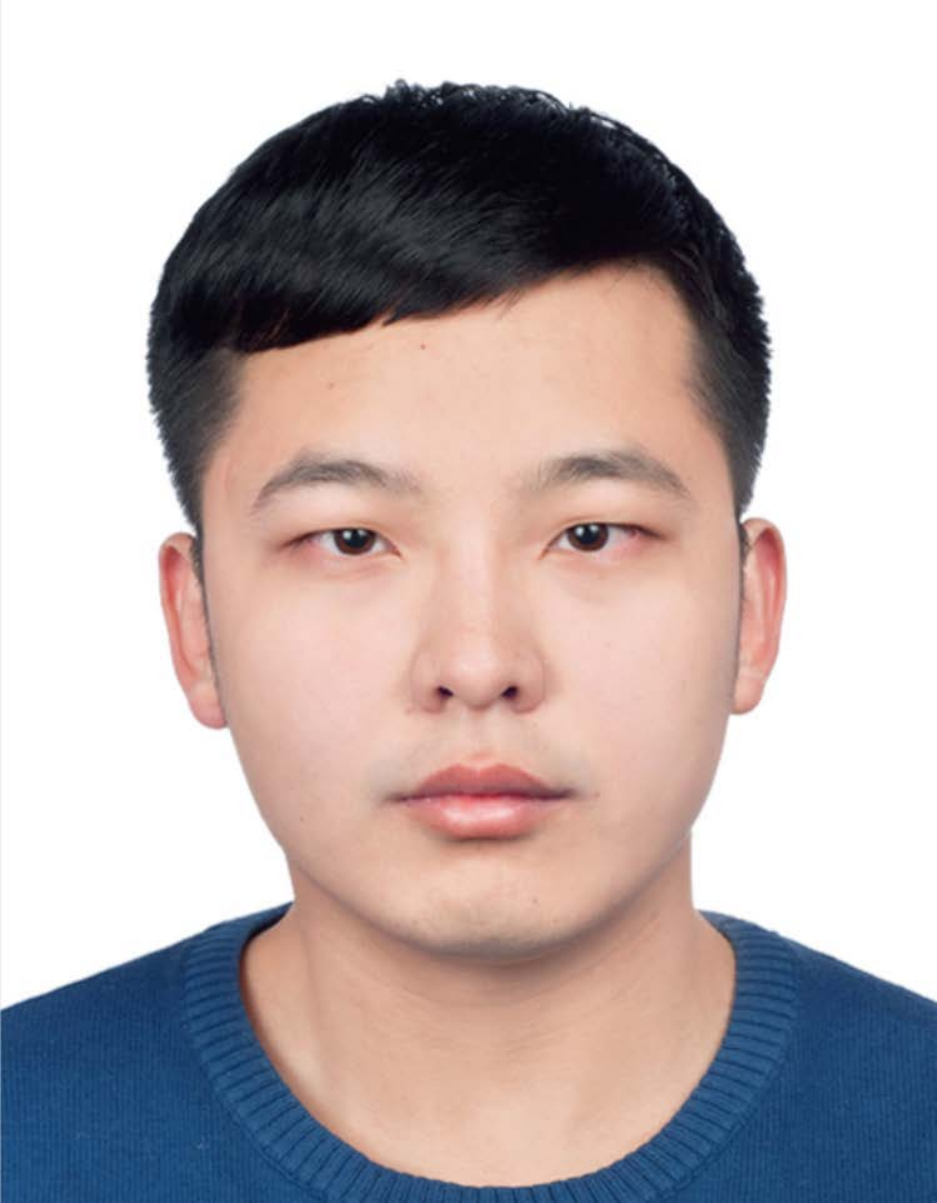}}]
    {Zhen~Zhuang}
    received his M.Eng. and B.Eng. from Fuzhou University in 2021 and 2018, respectively. He obtained his Ph.D. from the Chinese University of Hong Kong in 2025 under the supervision of Prof. Tsung-Yi Ho. His current research interest is Electronic Design Automation (EDA), especially EDA for advanced packaging and 3D IC. He was a recipient of three ICCAD/ISPD contest awards.
\end{IEEEbiography}
\vspace{-.2in}
\begin{IEEEbiography}
    [{\includegraphics[width=1.0in,height=1.26in,clip,keepaspectratio]{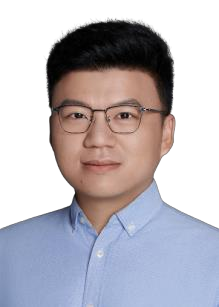}}]
    {Rongliang~Fu}
    received his Ph.D. in Computer Science and Engineering from The Chinese University of Hong Kong in January 2026, following an M.S. from the University of Chinese Academy of Sciences in June 2021 and a B.S. in Software Engineering from Northwestern Polytechnical University in June 2018. He has authored over 30 papers across major journals (IEEE TC and IEEE TCAD) and conferences (DAC, DATE, ICCAD, etc.). His research spans electronic design automation and EDA for superconducting electronics.
\end{IEEEbiography}
\vspace{-.4in}
\begin{IEEEbiography}
    [{\includegraphics[width=1.0in,height=1.26in,clip,keepaspectratio]{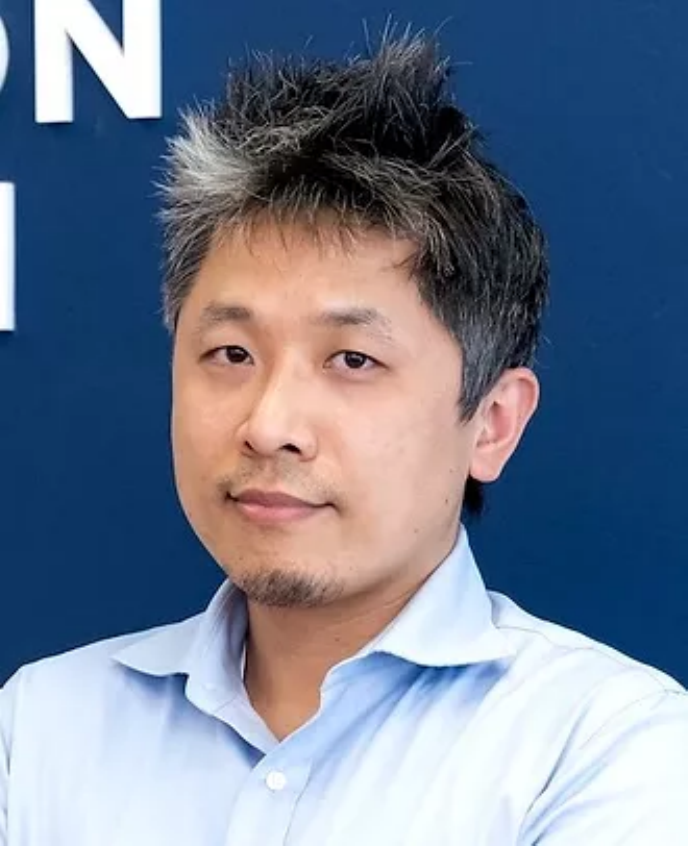}}]
    {Bei~Yu}
    (M'15-SM'22)
    received the Ph.D.~degree from The University of Texas at Austin in 2014.
    He is currently a Professor in the Department of Computer Science and Engineering, The Chinese University of Hong Kong.
    He has served as TPC Chair of ACM/IEEE Workshop on Machine Learning for CAD, and in many journal editorial boards and conference committees.
    He received ten Best Paper Awards from IEEE TSM 2022, DATE 2022, ICCAD 2021 \& 2013, ASPDAC 2021 \& 2012, ICTAI 2019, Integration, the VLSI Journal in 2018, ISPD 2017, SPIE Advanced Lithography Conference 2016, and many other awards, including DAC Under-40 Innovator Award (2024), IEEE CEDA Ernest S.~Kuh Early Career Award (2022), and Hong Kong RGC Research Fellowship Scheme (RFS) Award (2024).
\end{IEEEbiography}
\vspace{-.4in}
\begin{IEEEbiography}
    [{\includegraphics[width=1.0in,height=1.26in,clip,keepaspectratio]{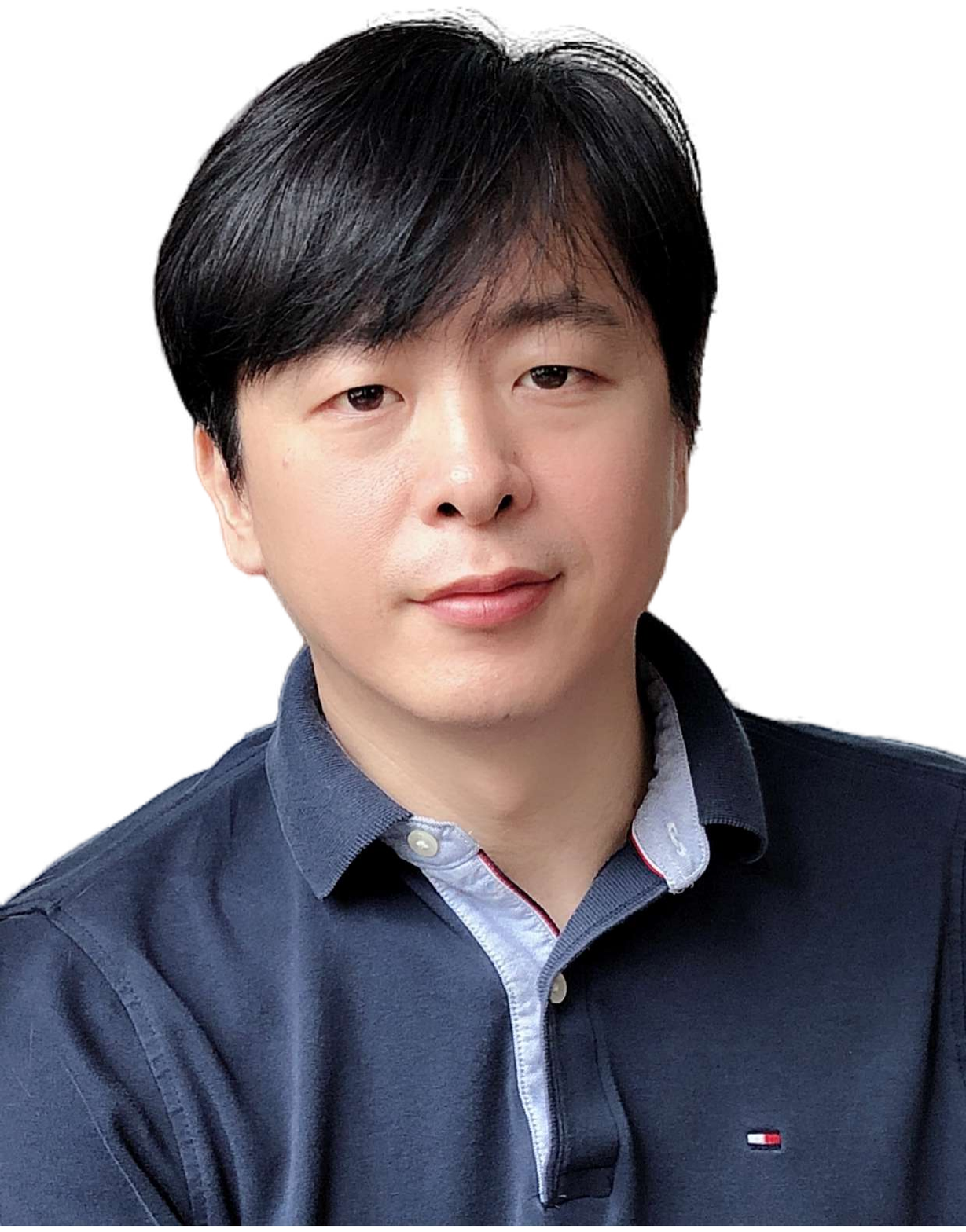}}]
    {Tsung-Yi~Ho}
    (F'24)
    is a Professor in the Department of Computer Science and Engineering, The Chinese University of Hong Kong (CUHK). He received his Ph.D. in Electrical Engineering from National Taiwan University in 2005. His research interests include several areas of computing and emerging technologies, especially in the design automation of microfluidic biochips. He was a recipient of the Best Paper Award at the IEEE Transactions on Computer-Aided Design of Integrated Circuits and Systems in 2015. Currently, he serves as the VP Conferences of IEEE CEDA, and the Executive Committee of ASP-DAC and ICCAD. He is a Distinguished Member of ACM and a Fellow of IEEE.
\end{IEEEbiography}

%% file: Clip3D.bbl
\begin{thebibliography}{10}
\providecommand{\url}[1]{#1}
\csname url@samestyle\endcsname
\providecommand{\newblock}{\relax}
\providecommand{\bibinfo}[2]{#2}
\providecommand{\BIBentrySTDinterwordspacing}{\spaceskip=0pt\relax}
\providecommand{\BIBentryALTinterwordstretchfactor}{4}
\providecommand{\BIBentryALTinterwordspacing}{\spaceskip=\fontdimen2\font plus
\BIBentryALTinterwordstretchfactor\fontdimen3\font minus \fontdimen4\font\relax}
\providecommand{\BIBforeignlanguage}[2]{{%
\expandafter\ifx\csname l@#1\endcsname\relax
\typeout{** WARNING: IEEEtran.bst: No hyphenation pattern has been}%
\typeout{** loaded for the language `#1'. Using the pattern for}%
\typeout{** the default language instead.}%
\else
\language=\csname l@#1\endcsname
\fi
#2}}
\providecommand{\BIBdecl}{\relax}
\BIBdecl

\bibitem{xing2024codesign3DIC}
D.~Xing \emph{et~al.}, ``A high level approach to co-designing 3d ics,'' in \emph{DAC}, 2024.

\bibitem{huang2024evaluation}
D.~Huang \emph{et~al.}, ``An evaluation framework for dynamic thermal management strategies in 3d multiprocessor system-on-chip co-design,'' \emph{IEEE Transactions on Parallel and Distributed Systems}, 2024.

\bibitem{gem5_simulator}
N.~Binkert \emph{et~al.}, ``The gem5 simulator,'' \emph{ACM SIGARCH computer architecture news}, 2011.

\bibitem{gem5_simulator_v20}
J.~Lowe-Power, A.~M. Ahmad, A.~Akram, M.~Alian, R.~Amslinger, M.~Andreozzi, A.~Armejach, N.~Asmussen, B.~Beckmann, S.~Bharadwaj \emph{et~al.}, ``The gem5 simulator: Version 20.0+,'' \emph{arXiv preprint arXiv:2007.03152}, 2020.

\bibitem{SimpleScalar_simulator}
D.~Burger and T.~M. Austin, ``The {SimpleScalar} tool set, version 2.0,'' \emph{SIGARCH Comput. Archit. News}, vol.~25, no.~3, pp. 13--25, 1997.

\bibitem{2004ICCAD_JasonCong_t3dflp}
J.~Cong, J.~Wei, and Y.~Zhang, ``A thermal-driven floorplanning algorithm for {3D ICs},'' in \emph{IEEE/ACM International Conference on Computer-Aided Design (ICCAD)}.\hskip 1em plus 0.5em minus 0.4em\relax IEEE, 2004, pp. 306--313.

\bibitem{2006_JasonCong_theraml_PD}
J.~Cong and Y.~Zhang, ``Thermal-aware physical design flow for {3-D ICs},'' in \emph{Proc. International VLSI Multilevel Interconnection Conference}, 2006, pp. 73--80.

\bibitem{2006ASPDAC_JasonCong_meva3d}
J.~Cong, A.~Jagannathan, Y.~Ma, G.~Reinman, J.~Wei, and Y.~Zhang, ``An automated design flow for {3D} microarchitecture evaluation,'' in \emph{IEEE/ACM Asia and South Pacific Design Automation Conference (ASPDAC)}, 2006, pp. 384--389.

\bibitem{2007ICCD_JasonCong_3d_exploration}
Y.~Liu, Y.~Ma, E.~Kursun, G.~Reinman, and J.~Cong, ``Fine grain {3D} integration for microarchitecture design through cube packing exploration,'' in \emph{IEEE/ACM International Conference on Computer-Aided Design (ICCAD)}.\hskip 1em plus 0.5em minus 0.4em\relax IEEE, 2007, pp. 259--266.

\bibitem{2021TVLSI-thermal-aware-3dfloorplan-tsv}
J.-M. Lin, W.-Y. Chang, H.-Y. Hsieh, Y.-T. Shyu, Y.-J. Chang, and J.-M. Lu, ``{Thermal-aware floorplanning and TSV-planning for mixed-type modules in a fixed-outline 3-D IC},'' \emph{IEEE Transactions on Very Large Scale Integration Systems (TVLSI)}, vol.~29, no.~9, pp. 1652--1664, 2021.

\bibitem{sa3dfp_guan_2023}
W.~Guan, X.~Tang, H.~Lu, Y.~Zhang, and Y.~Zhang, ``A novel thermal-aware floorplanning and {TSV} assignment with game theory for fixed-outline 3-{D} {ICs},'' \emph{IEEE Transactions on Very Large Scale Integration (VLSI) Systems}, vol.~31, no.~11, pp. 1639--1652, 2023.

\bibitem{wang2025cool3d}
R.~Wang, Z.~Wang, T.~Lin, J.~M. Raby, M.~R. Stan, and X.~Guo, ``{Cool-3D}: An end-to-end thermal-aware framework for early-phase design space exploration of microfluidic-cooled {3DICs},'' \emph{arXiv preprint arXiv:2503.07297}, 2025.

\bibitem{hankin2021hotgauge}
A.~Hankin, D.~Werner, M.~Amiraski, J.~Sebot, K.~Vaidyanathan, and M.~Hempstead, ``{HotGauge}: A methodology for characterizing advanced hotspots in modern and next generation processors,'' in \emph{IEEE International Symposium on Workload Characterization (IISWC)}.\hskip 1em plus 0.5em minus 0.4em\relax IEEE, 2021, pp. 163--175.

\bibitem{HotLEGO}
R.~Wang, J.-H. Han, M.~Stan, and X.~Guo, ``{Hot-LEGO}: Architect microfluidic cooling equipped {3DIC} with pre-{RTL} thermal simulation,'' in \emph{International Green and Sustainable Computing Conference}, 2023, pp. 14--17.

\bibitem{li2009mcpat}
S.~Li \emph{et~al.}, ``Mcpat: An integrated power, area, and timing modeling framework for multicore and manycore architectures,'' in \emph{MICRO}, 2009.

\bibitem{muralimanohar2007cacti}
N.~Muralimanohar, R.~Balasubramonian, and N.~P. Jouppi, ``Optimizing {NUCA} organizations and wiring alternatives for large caches with {CACTI 6.0},'' in \emph{IEEE/ACM International Symposium on Microarchitecture (MICRO)}.\hskip 1em plus 0.5em minus 0.4em\relax IEEE, 2007, pp. 3--14.

\bibitem{hotspot_stan2003}
M.~R. Stan \emph{et~al.}, ``Hotspot: A dynamic compact thermal model at the processor architecture level,'' \emph{Microelectronics Journal}, 2003.

\bibitem{hotspot6}
R.~Zhang, M.~R. Stan, and K.~Skadron, ``Hotspot 6.0: Validation, acceleration and extension,'' \emph{University of Virginia, Tech. Rep}, vol.~15, no.~4, pp. 1--8, 2015.

\bibitem{hotspot7}
J.-H. Han, R.~E. West, K.~Skadron, and M.~R. Stan, ``Thermal simulation of processing-in-memory devices using hotspot 7.0,'' in \emph{2021 27th international workshop on thermal investigations of ICs and systems (THERMINIC)}.\hskip 1em plus 0.5em minus 0.4em\relax IEEE, 2021, pp. 1--5.

\bibitem{ahmed2016tsv}
M.~A. Ahmed, S.~Mohapatra, and M.~Chrzanowska-Jeske, ``{TSV-} and delay-aware {3D-IC} floorplanning,'' \emph{Analog Integrated Circuits and Signal Processing}, vol.~87, no.~2, pp. 235--248, 2016.

\bibitem{2022ISLPED_lim_3dic_ppa_thermal}
L.~Zhu, N.~E. Bethur, Y.-C. Lu, Y.~Cho, Y.~Im, and S.~K. Lim, ``3d ic tier partitioning of memory macros: Ppa vs. thermal tradeoffs,'' in \emph{IEEE International Symposium on Low Power Electronics and Design (ISLPED)}, 2022, pp. 1--6.

\bibitem{Kendall_s_tau}
P.~K. Sen, ``Estimates of the regression coefficient based on kendall's tau,'' \emph{Journal of the American statistical association}, 1968.

\bibitem{Cool3D}
R.~Wang \emph{et~al.}, ``Cool-3d: An end-to-end thermal-aware framework for early-phase design space exploration of microfluidic-cooled 3dics,'' \emph{IEEE Journal on Emerging and Selected Topics in Circuits and Systems}, 2025.

\bibitem{lbfgs_proposed}
R.~H. Byrd, P.~Lu, J.~Nocedal, and C.~Zhu, ``A limited memory algorithm for bound constrained optimization,'' \emph{SIAM Journal on Scientific Computing}, vol.~16, no.~5, pp. 1190--1208, 1995.

\bibitem{wu2005joint_cyw}
Y.-W. Wu \emph{et~al.}, ``Joint exploration of architectural and physical design spaces with thermal consideration,'' in \emph{Proceedings of the 2005 international symposium on Low power electronics and design}, 2005.

\bibitem{dally2004principles}
W.~J. Dally and B.~P. Towles, \emph{Principles and Practices of Interconnection Networks}.\hskip 1em plus 0.5em minus 0.4em\relax Morgan Kaufmann, 2004.

\bibitem{chandrakasan1992lowpower}
A.~P. Chandrakasan, S.~Sheng, and R.~W. Brodersen, ``Low-power {CMOS} digital design,'' \emph{IEEE Journal of Solid-State Circuits}, vol.~27, no.~4, pp. 473--484, 1992.

\bibitem{1995ICCAD-FM_partition}
C.~M. Fiduccia and R.~M. Mattheyses, ``A linear-time heuristic for improving network partitions,'' in \emph{Proc.\ DAC}, 1982, pp. 175--181.

\bibitem{bakoglu1985optimal}
H.~Bakoglu and J.~D. Meindl, ``Optimal interconnection circuits for vlsi,'' \emph{IEEE Transactions on Electron Devices}, vol.~32, no.~5, pp. 903--909, 1985.

\bibitem{zhao2007predictive}
W.~Zhao and Y.~Cao, ``Predictive technology model for nano-cmos design exploration,'' \emph{ACM Journal on Emerging Technologies in Computing Systems (JETC)}, vol.~3, no.~1, pp. 1--es, 2007.

\bibitem{splash2_1995}
S.~C. Woo, M.~Ohara, E.~Torrie, J.~P. Singh, and A.~Gupta, ``The {SPLASH-2} programs: Characterization and methodological considerations,'' in \emph{IEEE/ACM International Symposium on Computer Architecture (ISCA)}, 1995, pp. 24--36.

\bibitem{mccalpin1995stream}
J.~D. McCalpin, ``Memory bandwidth and machine balance in current high performance computers,'' \emph{IEEE Computer Society Technical Committee on Computer Architecture (TCCA) Newsletter}, pp. 19--25, 1995.

\end{thebibliography}
